\documentclass[aps,prb,a4paper,reprint,nobibnotes,superscriptaddress]{revtex4-1}
\usepackage{graphicx,graphics}
\usepackage{amssymb,amsmath,amsfonts}
\usepackage{array}
\usepackage{dcolumn}
\usepackage{soul}
\usepackage{color}
\usepackage{natbib}
\usepackage[svgnames]{xcolor}
\usepackage{dcolumn}
\newcolumntype{d}{D{.}{\cdot}{2}}
\begin{document}

\title{Exact Wave Packet Dynamics of Singlet Fission in Unsubstituted and Substituted Polyene Chains 
within Long-Range Interacting Models}

\date{\today}

\author{Suryoday Prodhan}
\email[Electronic mail: ]{suryodayp@sscu.iisc.ernet.in}
\affiliation{Solid State and Structural Chemistry Unit, Indian Institute of Science,
Bangalore 560012, India}
\author{S. Ramasesha}
\email[Electronic mail: ]{ramasesh@sscu.iisc.ernet.in}
\affiliation{Solid State and Structural Chemistry Unit, Indian Institute of Science,
Bangalore 560012, India}

\begin{abstract}

Singlet fission (SF) is a potential pathway for significant enhancement of efficiency in organic solar cells 
(OSC). In this paper, we study singlet fission in a pair of polyene molecules in two different stacking 
arrangements employing exact many-body wave packet dynamics. In the non-interacting model, the SF yield is absent.
The individual molecules are treated within Hubbard and Pariser-Parr-Pople (PPP) models and the interaction 
between them involves transfer terms, intersite electron repulsions and site-charge--bond-charge repulsion terms. 
Initial wave packet is constructed from excited singlet state of one molecule and ground state of the other. 
Time development of this wave packet under the influence of intermolecular interactions is followed within the 
Schr\"odinger picture by an efficient predictor-corrector scheme. In unsubstituted Hubbard and PPP chains, 
$2{}^1A$ excited singlet state leads to significant SF yield while the $1{}^1B$ state gives negligible fission 
yield. On substitution by donor-acceptor groups of moderate strength, the lowest excited state will have 
sufficient $2{}^1A$ character and hence results in significant SF yield. 
Because of rapid internal conversion, the nature of the lowest excited singlet will determine the SF contribution 
to OSC efficiency.
Furthermore, we find the fission yield depends considerably on the stacking arrangement of the polyene molecules.

\end{abstract}



\maketitle

\section{Introduction}

Singlet fission (SF) is a process in which a molecule in the singlet excited state $(S_n)$ interacts with another 
molecule in the ground state $(S_0)$ resulting in triplet excited state on each molecule \cite{Schneider65,
*Swenberg68,*Merrifield69,*Merrifield70-1}. Although this process can be described by a single step reversible 
pathway, a detailed scheme \cite{Merrifield68,*Merrifield70-2} considering the hypotheses of spin-allowed 
transition is vividly accepted by the scientific community. In this scheme, interaction between the $S_n$ state 
and $S_0$ state results in a spin-singlet coupled ${}^1(T_1T_1)$ state which later dissociates into two triplet 
excitons (Ref. \citenum{Michl10,Michl13-1,Michl13-2} and references therein). Recently it has been reported that 
this multiexciton state is being observed experimentally via time-resolved two-photon photoemission spectroscopy 
{\cite{Chan11}} and transient absorption and time-resolved photoluminescence spectroscopy {\cite{Walker15}}. These
studies also have shown more than $100\%$ triplet yield from singlet excited state.
In promising systems the rate constants for the fission of the singlet excited state should be higher, compared 
to other intra- and intermolecular processes like fluorescence. The energetics for singlet fission consists of 
two widely accepted requirements{\cite{Michl10}}-- (i) $E_{S_n}\geq2E_{T_1}$; systems with 
$E_{S_n}$ slightly less than $2E_{T_1}$, have also been found to display signature of singlet fission due to 
vibronic processes (ii) the energy of the higher triplet state $(T_2)$ should be greater than $S_n$ energy to 
prevent intersystem crossing to the triplet state and also should be more than twice of $T_1$ energy to suppress 
re-fusion of the newly born triplets by triplet-triplet annihilation; however, in finite-size polyene 
chains $T_2$ remain lower compared to $S_n$, irrespective of the symmetry of the $S_n$ state and the second 
criteria is not met. In literature, $S_1$ which is the lowest singlet excited state is commonly considered to be 
the optically excited state. Michl and coworkers proposed a number of suitable SF candidates which satisfy the 
energetics criteria on the basis of single CI calculations within Pariser-Parr-Pople model \cite{Michl06} and 
speculated that alternant hydrocarbons (notably polyacenes) and biradicaloids are good choices as chromophores for
SF. In another study, Greyson and coworkers examined the appropriate strength of interchromophoric coupling 
necessary for singlet fission in some promising materials, employing density functional theory (DFT) 
{\cite{Greyson10-2}}. Minami and Nakano gave a biradical description of singlet fission considering biradicaloid 
systems which have open-shell ground states \cite{Nakano12-1}. They have also studied small-size oligorylenes 
{\cite{Nakano12-2}} and alternant and non-alternant hydrocarbons {\cite{Nakano13}} as singlet fission candidates 
employing the time-dependent density functional theory (TD-DFT). However, the notion that both lowest singlet 
excited state and lowest triplet state can be described by HOMO-LUMO excitations from the ground state is too 
crude for $\pi$-conjugated systems.

In this paper, we have gone beyond the static quantum chemical approach and studied the quantum dynamics of 
singlet fission. We have considered dimers of 1,3-butadiene, 1,3,5-hexatriene and 1,3,5,7-octatetraene in full 
configuration interaction space of the $\pi-$system within Hubbard and Pariser-Parr-Pople (PPP) model 
Hamiltonians. In the literature, there exist only a few studies which go beyond frontier molecular orbitals 
approximation \cite{Zimmerman10,Zimmerman11,Zimmerman12,Zeng14,Ananth14,Ananth16,deGier12,Parker14,Aryanpour15-1}.
The polyene systems are important model molecules and there are several reports which indicate singlet fission in 
carotenoids and polyene systems {\cite{Aryanpour15-1,Gradinaru01,Kennis02,Tauber10,Tauber12,Bardeen13-2,Musser13,
Musser15}}. We start with a wave packet formed from the ground state of one molecule and the singlet excited state of 
another, these states being the exact eigenstates of the molecule within the chosen model Hamiltonians. We then 
introduce intermolecular interactions and evolve the wave packet in time. At each time step, the evolved wave 
packet is projected on to various direct products of the eigenstates of individual molecules in the triplet 
manifold to obtain the yield of the triplets. The time evolution is carried out in the full configuration space 
of the total $\pi-$system.

Effect of crystal stacking on SF efficiency have been studied extensively for acenes and other 
hydrocarbons {\cite{Krylov13,Ratner13,Wasielewski13,Beljonne14,Grozema15}}. In most of the materials, slipped stacked 
arrangement results in higher singlet fission yield as intermolecular vibrational modes which lead to direct 
coupling between the $S_0S_1$ state and ${}^1(T_1T_1)$ state are sensitive to crystal packing. 
However, reports by Friend et al. and Guldi et al. on solution phase SF for substituted pentacene pointed out 
that SF is not confined to specific geometries and can be observed even in disordered systems {\cite{Friend13,
Guldi15}}. Similar conclusion is also arrived at by Sanders and coworkers who studied SF of bipentacene in solution 
phase {\cite{Steigerwald15}}.

The $S_1$ states in these polyenes are optically inactive and are primarily composed of two triplets 
\cite{Hudson72,*Schulten72,Schulten87}. Substitution in these moieties by donor-acceptor groups breaks the 
electron-hole and inversion symmetries making $S_1$ state optically active and therefore the lowest optical state 
in these systems shifts from $S_2$ to $S_1$. However, for weak symmetry breaking $S_1$ state continues to show 
the characteristics of two triplets. Thus we find that even in substituted polyenes, if the initial state is an 
$S_1$ state rather than other higher energy singlet state, SF is efficient. This agrees with 
some recent studies which suggest substitution by heteroatoms within organic chromophore {\cite{Greyson10-1,
Zeng14,Chen14}} or copolymerising donor-acceptor moieties {\cite{Sfeir15,Aryanpour15-2,Kasai15}} can play an 
important role in tailoring candidate molecules for SF. The mixing of different eigenstates on donor-acceptor 
substitution, although well-known in $\pi$-conjugated carbon systems, its significance in singlet fission has 
not been explored. Our model study will be helpful in providing insights for developing better systems for SF.

In recent years, there is also considerable interest in intramolecular SF (iSF) in polymers. 
There are primarily two classes of systems which have been widely studied. In one class, the polymers consists 
of chromophores linked via conjugated linkers or covalent linkers which lead to through-bond or through-space 
interactions respectively. The widely studied systems have primarily polyacene chromophores like tetracene 
{\cite{Bardeen06,Bardeen07,Damrauer13,Thompson16,Krylov16,Damrauer16}}, pentacene {\cite{Ananth16,Guldi15,
Steigerwald15,Greenham15,Sfeir16-1,Sfeir16-2,Hasobe16,Nakano16,Guldi16,Musser16,Sfeir16-3}}, 
1,3-diphenylisobenzofuran {\cite{Michl13-3}} or terrylenediimide {\cite{Wasielewski16}}. iSF studies have also 
been reported on bithiophene~{\cite{Goodson15,Zimmerman15}} and P3TV polymer~{\cite{Musser13}} which belong to 
the same class. The second class of systems consist of strong donor-acceptor units in the polymer which act as 
chromophores. Some notable candidates belonging to this class of iSF systems are PBTDO1 and PBTDO2 
~{\cite{Sfeir15}}, PDTP-DFBT~{\cite{Aryanpour15-2}} and PTB1~{\cite{Kasai15}}. In our study, chromophores in the 
respective systems are two polyene chains which have through-space interactions due to molecular stacking. Hence, 
our study can also be viewed as intramolecular SF with through-space interactions between chromophores. Since, 
the difference between intramolecular and intermolecular SF is more semantic 
than substantial, a time evolution study of intramolecular SF will also proceed along similar lines as our study.

In the following section, we have given a brief account of the model Hamiltonians and methodology used in our 
study. In section III, we have discussed the pictures which emerge for unsubstituted polyenes 
within different model Hamiltonians along with the role of substitution in singlet fission yield for different 
alkene chain. In section IV, we summarize our study.

\section{Methodology}

In our study, the individual molecules considered are polyenes which have chain lengths, $N$ varying from 
$4$ to $8$ sites and are modelled by the Pariser-Parr-Pople (PPP) Hamiltonian \cite{pariser-parr,pople}, 
which includes long-range electron correlations along with on-site Hubbard interaction $(U)$. 
The Hamiltonian of individual polyene is given by:

\begin{equation}
\begin{split}
H_{intra} &= \sum_{i=1}^{N-1} t_0(1-(-1)^i \delta)(\hat E_{i,i+1}+\mbox{ H.C.})+ 
\sum_{i=1}^N \epsilon_i \hat n_i \\ 
&+\sum_{i=1}^N \frac {U}{2} \hat n_i (\hat n_i-1) 
+ \sum_{i>j=1}^N V_{ij}(\hat n_i-z_i)(\hat n_j-z_j) 
\end{split}
\label{ppp-intra}
\end{equation}
\begin{equation*}
\hat E_{i,i+1}= \sum_{\sigma} \hat c_{i,\sigma}^\dagger \hat c_{i+1,\sigma}^{}
\end{equation*}
\noindent
where $t_0$ is the average transfer integral; $\delta$ is the strength of dimerization; $\epsilon_i$ is the 
site energy at the $i$-th site; $U$ is the Hubbard correlation strength and $V_{ij}$-s are the intersite 
electronic correlation strengths; $z_i$ is the local chemical potential at site $i$ which leaves the site 
neutral (for carbon in a $\pi$-conjugated system $z_i=1$). $\hat c_{i,\sigma}^\dagger$ $(\hat c_{i,\sigma}^{})$ 
creates (annihilates) an electron of spin $\sigma$ in the orbital at $i$-th site and $\hat n_i$ is the 
corresponding number operator. Standard PPP parameters for carbon are employed, namely $t_0=-2.40$ eV and 
$U=11.26$ eV. $\delta$ is taken as $0.07$ and the C-C bond lengths are fixed at 
$1.40(1+\frac {\delta}{2})$ \AA ~for the single bond and $1.40(1-\frac {\delta}{2})$ \AA ~for the double bond.
The long-range Coulomb interaction term $V_{ij}$ between site `i' and `j' is parameterized using Ohno 
interpolation scheme \cite{ohno,*klopman},
\begin{equation}
V_{ij}=14.397 \left[\left(\frac{14.397}{U}\right)^2+r_{ij}^2\right]^{-\frac{1}{2}}
\label{eqohno}
\end{equation}
which is arrived at by interpolating between $U$ at $r_{ij}=0$ and $e^2/r_{ij}$ for $r_{ij}\rightarrow$ $\infty$. 
In Eq. \ref{eqohno}, distance between site `i' and `j'~$(r_{ij})$ is in {\AA} while the energies are in 
eV ~\cite{race01,*race03}. To study the role of substitution, site energies are varied at the chain ends to mimic 
donor and acceptor groups. Positive site energies correspond to donor groups and negative site energies 
to acceptor groups while site energies of unsubstituted carbon atoms are all set to zero. 
In our study, we have varied the strength of donor-acceptor substitution, $|\epsilon|$, from $0$ to $5$ eV. 

\begin{figure}[t]
\begin{center}
\includegraphics[height=7.5cm,width=7.5cm]{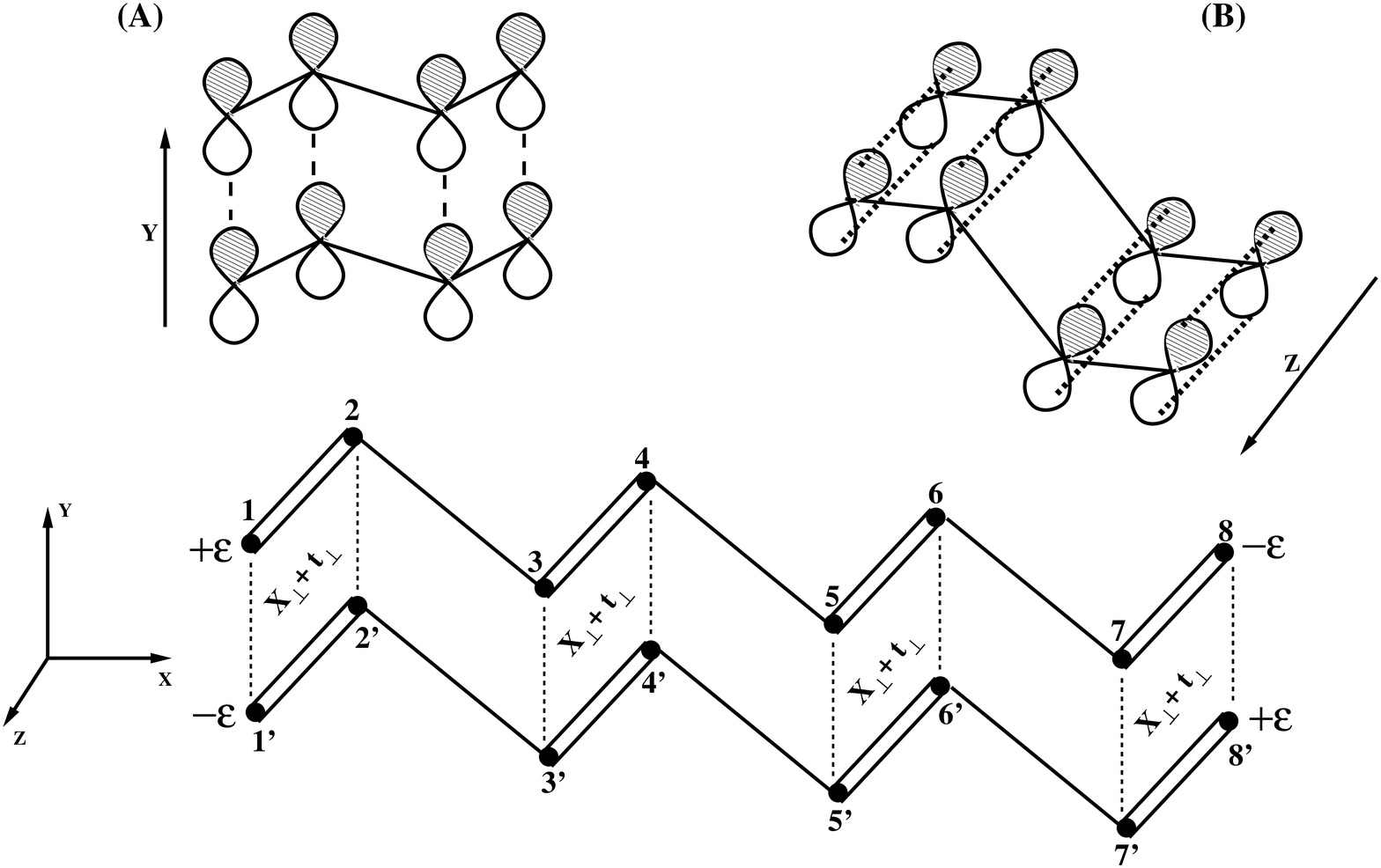}
\caption{\footnotesize{Schematic diagram of the stacked polyenes; (A) in vertical stacking, monomers 
get stacked along Y-axis; (B) in horizontal stacking, monomers get stacked along Z-axis.
XY plane is the molecular plane. The broken lines represent the intermolecular hopping interaction. 
$+\epsilon$ and $-\epsilon$ represent the donor and acceptor sites respectively while $\delta$ is the 
dimerization factor. The intrachain transfer integrals are taken to be $t_0(1\pm \delta)$ for 
double/single bonds and corresponding bond lengths are taken to be $r_0(1\mp \delta/2)$. $t_0$ and $r_0$
are chosen to be $2.40$ eV and $1.4$ \AA ~respectively. Site indices on different molecules are differentiated 
by using `prime' superscript for sites on one molecule and without `prime' for the other molecule.}}
\label{system}
\end{center}
\end{figure}

If all long-range intersite interaction terms in the Hamiltonian are discarded, it represents Hubbard Hamiltonian. 
Singlet fission in unsubstituted systems are also studied within this model Hamiltonian, as a function of 
$U/t_0$ to probe the role of correlation strength. The $U/t_0=0$ case will reproduce the non-interacting or 
H\"uckel picture. In both H\"uckel and Hubbard models, we have considered $t_0=-1.0$ eV and the dimerization 
strength same as in the PPP model.

\begin{figure*}[t]
\centering
\begin{minipage}[t][3cm][t]{0.45\textwidth}
\centering
\includegraphics[width=\textwidth,height=0.6\textwidth]{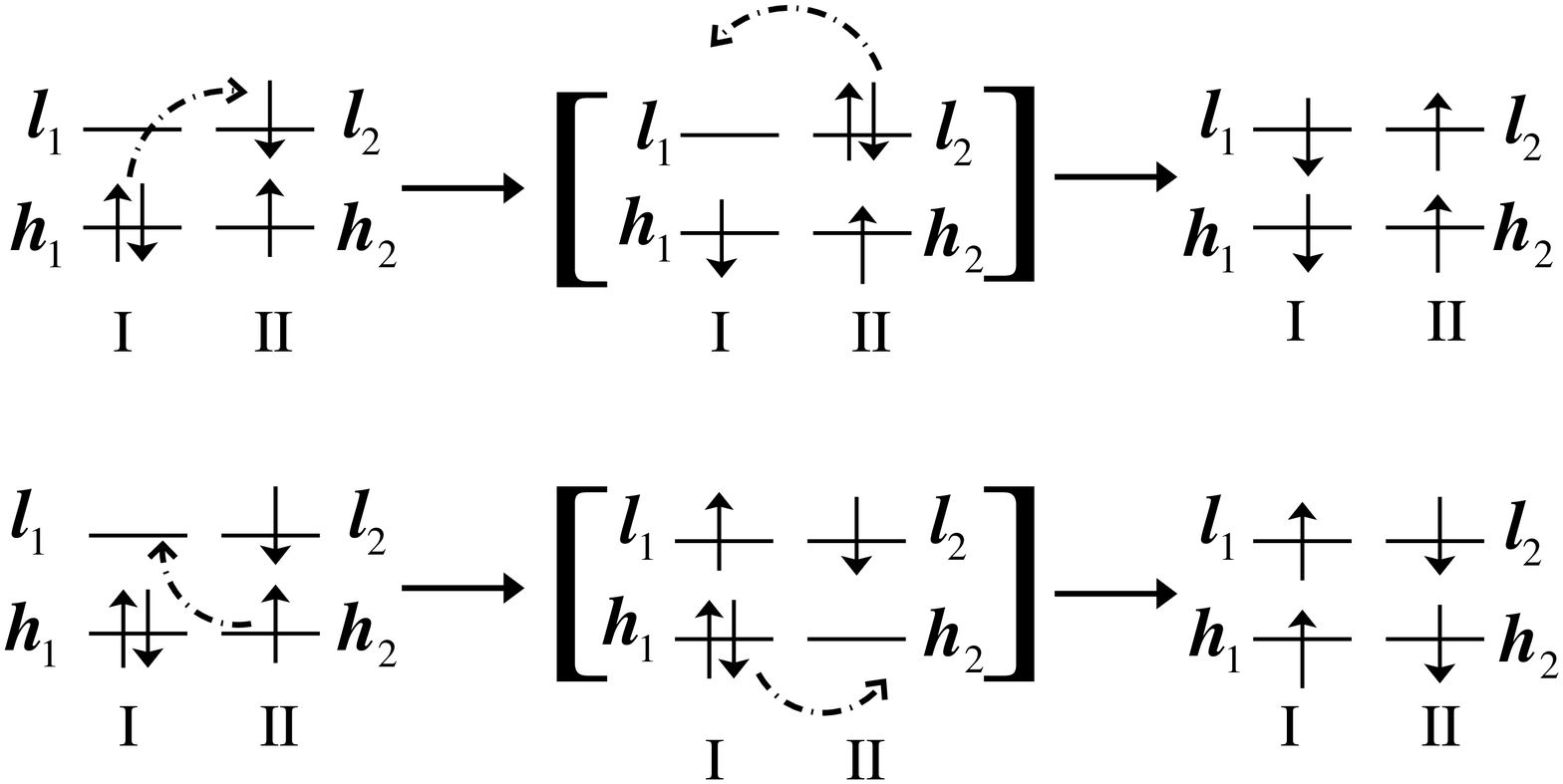}
(A)
\end{minipage}
\quad\quad\quad\quad\quad
\begin{minipage}[t][3cm][t]{0.35\textwidth}
\centering
\includegraphics[width=\textwidth,height=0.4\textwidth]{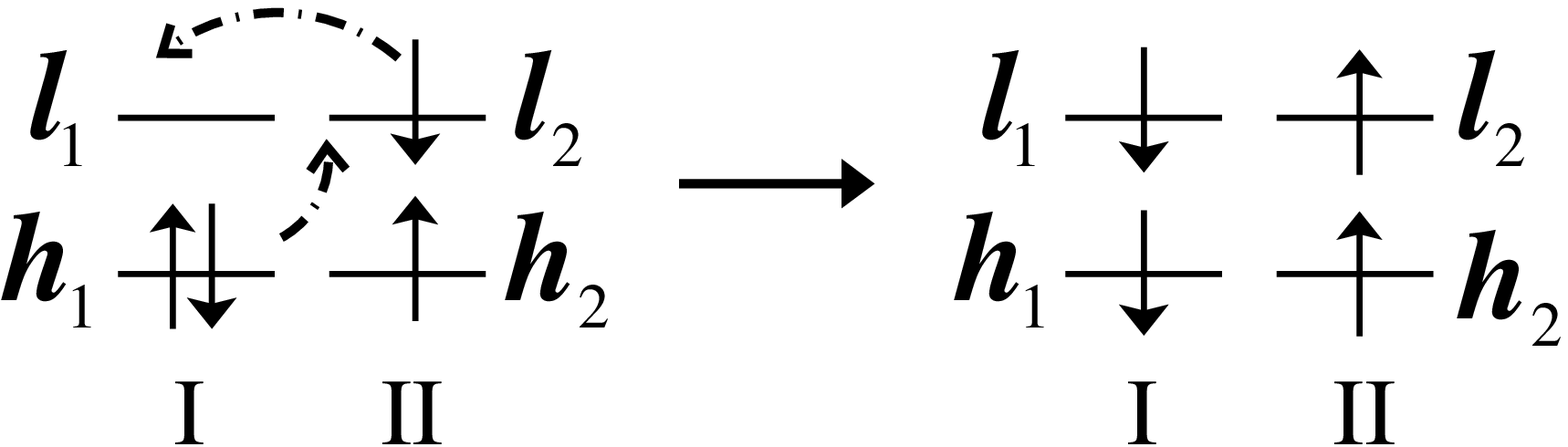}
(B)
\end{minipage}
\caption{\footnotesize{(A) Schematic diagram of singlet fission pathway involving two inter-chromophoric 
one-electron hopping interactions ($t_\perp$). 
$h_1$ and $l_1$ are the HOMO and LUMO of molecule $1$ while $h_2$ and $l_2$ are those of molecule $2$. 
The Hermitian conjugate of each step can be represented by reversing the direction of the broken arrow.
(B) Schematic diagram of another pathway via two-electron repulsion integral ($X_\perp$) involving frontier 
molecular orbitals.
The two-electron operators are $[l_1l_2|l_2h_1](\hat E_{l_1l_2} \hat E_{l_2h_1}- \hat E_{l_1h_1})$,
$[l_1l_2|h_1l_2]\hat E_{l_1l_2} \hat E_{h_1l_2}$, $[l_2l_1|l_2h_1]\hat E_{l_2l_1} \hat E_{l_2h_1}$,
$[l_2l_1|h_1l_2]\hat E_{l_2l_1} \hat E_{h_1l_2}$ and their Hermitian conjugates, where $\hat E_{ij}$ represents
$\sum_{\sigma} \hat a_{i,\sigma}^\dagger \hat a_{j,\sigma}$.}}
\label{mech}
\end{figure*}
 
The above Hamiltonians, being non-relativistic conserves total spin $S_{total}$, along with z-component of 
total spin $(S_{z,total})$. As we are primarily concerned with singlet and triplet manifolds, we work with 
valence bond (VB) basis which are eigenstates of total spin and employ the diagrammatic valence bond (DVB) 
method \cite{Soos84,Ramasesh84} for obtaining eigenstates in different spin subspaces 
for the monomers. Though complete and linearly independent, these basis states are non-orthogonal and result in 
non-symmetric sparse Hamiltonian matrices for the polyenes in question. The Hamiltonians are fully diagonalized 
in each case to obtain the complete spectrum within the singlet and triplet subspaces of individual polyenes.

For probing singlet fission, we have considered two polyene monomers arranged in an eclipsed conformation 
with the separation between the two set at $4$ \AA. The stacking orientation of the two monomers 
can be either ``vertical'' (V stacking), where one monomer remains on top of another or ``horizontal'' 
(H stacking), as shown in Fig. \ref{system}(A) and \ref{system}(B). In both orientations, these monomers remain 
in an electrostatically favorable stacking configuration where the donor (acceptor) site of molecule I lies 
directly above the acceptor (donor) site of molecule II (Fig. \ref{system}).

The intermolecular Hamiltonian between the two monomers is given by:
\begin{equation}
\begin{split}
H_{inter} &= \sum_{\langle i,i' \rangle} t_{\perp} (\hat E_{i,i'}+\hat E_{i',i}) \\ &+
\sum_{i} \sum_{j'} V_{ij'} (\hat n_i-z_i)(\hat n_{j'}-z_{j'}) \\ &+
\sum_{\langle i,i' \rangle} X_{\perp} (2\hat n_i+2\hat n_{i'}-2) (\hat E_{i,i'}+\hat E_{i',i})
\end{split}
\label{ppp-inter}
\end{equation}
\noindent
where $t_\perp$ is the inter-polyene hopping term between corresponding sites $i$ and $i'$ on chains I and II
which are directly above each other (Fig. \ref{system}). The transfer term is negative for horizontal stacking 
while it is positive for vertical stacking due to opposite signs of the overlap integrals (Fig. \ref{system}); 
in our calculations, we have considered $|t_\perp|=0.25$ eV in the PPP model and $0.2$ eV within 
H\"uckel and Hubbard models. The electron repulsion term comparable to the inter-molecular transfer term is 
$X_\perp$, the site-charge--bond-charge repulsion term and represents the two-electron integral $[ii|ii']$ and 
other related integrals within the charge cloud notation (Ref. \citenum{Campbell90,*Ramasesh03}); the other 
relevant multielectron repulsion term, bond-charge--bond-charge repulsion, represented by $[ii'|ii']$, is 
neglected as it is expected to be much smaller compared to $X_\perp$ (Ref. \citenum{Wolfsberg52,*Hoffmann62}). 
The site-charge--bond-charge term is neglected in the intra-molecular Hamiltonian as it affects only weakly the 
excitation spectrum of the isolated molecule. We have also taken $X_\perp=0$ for pair of sites on the two 
molecules which are not directly above each other. In charge cloud notation, the contribution to the Hamiltonian 
due to the repulsion term between site-charge at $i$ (of chain I) and the bond between $i$ and $i'$ (of chain II) 
can be denoted by the parameter $X_{\perp,\langle ii,ii' \rangle}=[ii|ii']+[ii|i'i]+[ii'|ii]+[i'i|ii]$; all 
integrals in this expression are equal and the corresponding second quantized operators are 
$(\hat n_i-1)\hat E_{ii'}$, $\hat n_i\hat E_{i'i}$, $\hat n_i\hat E_{ii'}$ and $(\hat n_i-1)\hat E_{i'i}$ 
respectively. Equivalent repulsion parameter $X_{\perp,\langle i'i',ii' \rangle}$ will also generate four 
interaction terms in the intermolecular Hamiltonian. In the non-interacting (H\"uckel) picture, $X_\perp$ term is 
taken to be zero. The pathways which lead to SF products from the $t_\perp$ and $X_\perp$ terms are schematically 
shown in Fig. \ref{mech}

To justify our treatment, it is important to show that the intermolecular interactions are weak enough to 
describe the state of the full system approximately by a product of the eigenstates of the individual polyenes. To 
demonstrate this, we have computed several properties in the low-lying eigenstates of the full system and compared them 
with those of isolated polyenes (Table. \ref{buta-1st} and Table. S1-S6 in the supplemental material \cite{supple}). These properties 
are (i) the projection of the low-lying eigenstates the full system onto 
the direct product of the low-lying eigenstates of isolated molecules (ii) average double occupancy of the sites in the 
monomer, and (iii) the spin of the monomer block in the full system calculated from the expectation value of the total 
spin $(S^2)$ for the block, using spin-spin correlation functions.

The ground state of the full system always has a very large projection on to the direct product of the ground 
states, namely, $S_0 \otimes S_0$. The double occupancy in the fragments is also the same as in the isolated molecules. 
The total spin of the fragments is also nearly zero. In the excited states there is always a singlet in the covalent space 
which has large projection to both the $T_1 \otimes T_1$ state and the $S_n \otimes S_0$ state. However, the average double 
occupancy of the sites from the two monomer states are very nearly the same as that of the full system. In this case, the spin expectation 
value on the fragments is large. In the case of other singlets in the ionic space, the projection on to the direct product 
of the isolated monomer states is very large. This analysis shows that the interaction term is a small perturbation on the 
isolated molecules.

Dynamics of a wave packet, which is direct product of a specified excited singlet state $|S_n^I\rangle$
of monomer I and the ground state $|S_0^{II}\rangle$ of monomer II, is studied in the Schr\"{o}dinger picture 
employing the full system Hamiltonian $H_{full}=H_{intra}+ H_{inter}$. We have chosen the Schr\"{o}dinger picture 
over the interaction picture as the full space of the dimer is too large to study within the interaction picture. 
It is convenient to obtain the eigenstates (both singlet and triplet) of the isolated molecules in the VB basis 
and convert them into Slater basis. The Hamiltonian matrix corresponding to $H_{full}$ is generated in the 
Hilbert space with $S_z^{full}=0$ using Slater basis. The wavepacket is time evolved employing the fourth-order 
multistep differencing scheme (MSD4)\cite{Iitaka94}, given by:

\begin{equation}
\begin{split}
|\psi&(t+2\Delta t)\rangle = |\psi(t-2\Delta t)\rangle + \frac{4iH_{full}\Delta t}{3}
\Biglb[|\psi(t)\rangle \Bigrb.\\&\Biglb. -2\biglb(|\psi(t+\Delta t)\rangle  
+|\psi(t-\Delta t)\rangle\bigrb) \Bigrb]+ O((H_{full}\Delta t)^5)
\label{msd4}
\end{split}
\end{equation}
\noindent
as the predictor and the fourth-order Adams-Moultan scheme (Eq. \ref{AM4}) as the corrector.
 
\begin{table*}[tbp]
\begin{center}
\caption{\label{buta-1st}\footnotesize{Properties of low-lying dimer eigenstates of unsubstituted 
butadiene and hexatriene dimers. $\langle\Psi|\Phi\rangle$ are projections of the dimer state on the direct 
product of the monomer states $|\Phi\rangle$, in the PPP model. 
The state $|\Phi\rangle$ is a simple direct product of the states when the states on the two monomers are the 
same. For two different monomer states, $|\phi\rangle$ and $|\chi\rangle$, $|\Phi\rangle = 
\frac{1}{\sqrt 2} (|\phi\rangle_I \otimes |\chi\rangle_{II} + |\chi\rangle_I \otimes |\phi\rangle_{II})$. 
$d_I$ and $d_{II}$ are average double occupancy per site within monomer I and  II while $s_I$ and $s_{II}$ are 
the spin value of corresponding monomer units calculated from the spin-spin correlation functions. 
$d_{_{I/II}}^{(iso)}$ are the average double occupancies of isolated monomer eigenstates. $\phi$-s and 
$\Psi$ are the eigenstates of the monomer and the dimer respectively.}}
\begin{ruledtabular}
\renewcommand{\arraystretch}{1.2}
\begin{tabular}{cccccccccccc}
N & Orientation & $\Psi$ & $\phi_{_I}$ & $\phi_{_{II}}$ & $|\langle \Psi|\Phi\rangle|$ & 
$d_{_I}^{(iso)}$ & $d_{_{II}}^{(iso)}$ & $d_{_I}$ & $d_{_{II}}$ & $s_{_I}$ & $s_{_{II}}$ \\
\colrule
4 & V & $S_0$ & $S_0$ & $S_0$ & $0.99$ & $0.17$ & $0.17$ & $0.17$ & $0.17$ & $0.004$ & $0.004$ \\
\cline{3-12}
  &   & $S_1$ & $S_1$ & $S_0$ & $0.44$ & $0.10$ & $0.17$ & $0.13$ & $0.13$ & $0.81$ & $0.81$ \\
\cline{4-8}
  &   &       & $T_1$ & $T_1$ & $0.83$ & $0.11$ & $0.11$ &        &        &        &        \\
\cline{3-12}
  &   & $S_2$ & $S_2$ & $S_0$ & $0.86$ & $0.31$ & $0.17$ & $0.24$ & $0.24$ & $0.15$ & $0.15$ \\
\cline{3-12}
  &   & $S_3$ & $S_1$ & $S_0$ & $0.95$ & $0.10$ & $0.17$ & $0.14$ & $0.14$ & $0.06$ & $0.06$ \\
\cline{3-12}
  &   & $S_4$ & $S_1$ & $S_0$ & $0.88$ & $0.10$ & $0.17$ & $0.13$ & $0.13$ & $0.33$ & $0.33$ \\
\cline{4-8}
  &   &       & $T_1$ & $T_1$ & $0.46$ & $0.11$ & $0.11$ &        &        &        &        \\
\cline{3-12}
  &   & $S_5$ & $S_2$ & $S_0$ & $0.96$ & $0.31$ & $0.17$ & $0.24$ & $0.24$ & $0.05$ & $0.05$ \\

\colrule

  & H & $S_0$ & $S_0$ & $S_0$ & $0.99$ & $0.17$ & $0.17$ & $0.17$ & $0.17$ & $0.003$ & $0.003$ \\
\cline{3-12}
  &   & $S_1$ & $S_1$ & $S_0$ & $0.35$ & $0.10$ & $0.17$ & $0.12$ & $0.12$ & $0.90$ & $0.90$ \\
\cline{4-8}
  &   &       & $T_1$ & $T_1$ & $0.92$ & $0.11$ & $0.11$ &        &        &        &        \\
\cline{3-12}
  &   & $S_2$ & $S_1$ & $S_0$ & $0.99$ & $0.10$ & $0.17$ & $0.14$ & $0.14$ & $0.01$ & $0.01$ \\
\cline{3-12}
  &   & $S_3$ & $S_1$ & $S_0$ & $0.93$ & $0.10$ & $0.17$ & $0.14$ & $0.14$ & $0.22$ & $0.22$ \\
\cline{4-8}
  &   &       & $T_1$ & $T_1$ & $0.36$ & $0.11$ & $0.11$ &        &        &        &        \\
\cline{3-12}
  &   & $S_4$ & $S_2$ & $S_0$ & $0.98$ & $0.31$ & $0.17$ & $0.24$ & $0.24$ & $0.04$ & $0.04$ \\
\cline{3-12}
  &   & $S_5$ & $S_2$ & $S_0$ & $0.98$ & $0.31$ & $0.17$ & $0.24$ & $0.24$ & $0.04$ & $0.04$ \\

\colrule
\colrule

6 & V & $S_0$ & $S_0$ & $S_0$ & $0.99$ & $0.18$ & $0.18$ & $0.18$ & $0.18$ & $0.007$ & $0.007$ \\
\cline{3-12}

  &   & $S_1$ & $S_1$ & $S_0$ & $0.45$ & $0.14$ & $0.18$ & $0.16$ & $0.16$ & $0.78$ & $0.78$ \\
\cline{4-8}
  &   &       & $T_1$ & $T_1$ & $0.80$ & $0.14$ & $0.14$ &        &        &        &        \\
\cline{3-12}
  &   & $S_2$ & $S_1$ & $S_0$ & $0.95$ & $0.14$ & $0.18$ & $0.16$ & $0.16$ & $0.07$ & $0.07$ \\
\cline{3-12}
  &   & $S_3$ & $S_2$ & $S_0$ & $0.83$ & $0.26$ & $0.18$ & $0.22$ & $0.22$ & $0.18$ & $0.18$ \\
\cline{3-12}
  &   & $S_4$ & $S_1$ & $S_0$ & $0.86$ & $0.14$ & $0.18$ & $0.15$ & $0.15$ & $0.36$ & $0.36$ \\
\cline{4-8}
  &   &       & $T_1$ & $T_1$ & $0.48$ & $0.14$ & $0.14$ &        &        &        &        \\
\cline{3-12}
  &   & $S_7$ & $S_2$ & $S_0$ & $0.95$ & $0.26$ & $0.18$ & $0.22$ & $0.22$ & $0.06$ & $0.06$ \\

\colrule

  & H & $S_0$ & $S_0$ & $S_0$ & $0.99$ & $0.18$ & $0.18$ & $0.18$ & $0.18$ & $0.004$ & $0.004$ \\ 
\cline{3-12}
  &   & $S_1$ & $S_1$ & $S_0$ & $0.41$ & $0.14$ & $0.18$ & $0.15$ & $0.15$ & $0.87$ & $0.87$ \\
\cline{4-8}
  &   &       & $T_1$ & $T_1$ & $0.90$ & $0.14$ & $0.14$ &        &        &        &        \\
\cline{3-12}
  &   & $S_2$ & $S_1$ & $S_0$ & $0.99$ & $0.14$ & $0.18$ & $0.16$ & $0.16$ & $0.01$ & $0.01$ \\
\cline{3-12}
  &   & $S_3$ & $S_1$ & $S_0$ & $0.90$ & $0.14$ & $0.18$ & $0.16$ & $0.16$ & $0.28$ & $0.28$ \\
\cline{4-8}
  &   &       & $T_1$ & $T_1$ & $0.42$ & $0.14$ & $0.14$ &        &        &        &        \\
\cline{3-12}
  &   & $S_4$ & $S_2$ & $S_0$ & $0.96$ & $0.26$ & $0.18$ & $0.22$ & $0.22$ & $0.06$ & $0.06$ \\
\cline{3-12}
  &   & $S_5$ & $S_2$ & $S_0$ & $0.96$ & $0.26$ & $0.18$ & $0.22$ & $0.22$ & $0.05$ & $0.05$ \\

\end{tabular}
\end{ruledtabular}
\end{center}
\end{table*}

\begin{equation}
\begin{split}
|\psi(t+  2\Delta t)&\rangle = |\psi(t+\Delta t)\rangle - \frac{iH_{full}\Delta t}{24}\Biglb[ 
9|\psi(t+2\Delta t)\rangle \Bigrb.\\&\Biglb. +19|\psi(t+\Delta t)\rangle -5|\psi(t)\rangle 
+ |\psi(t-\Delta t)\rangle \Bigrb]
\end{split}
\label{AM4}
\end{equation}
\noindent
This predictor-corrector scheme \cite{Dutta07,Chapra} is found to be very robust with accuracy comparable to the 
unconditionally stable Crank-Nicholson (CN) scheme \cite{Crank-Nicolson47}.
\begin{equation}
\begin{split}
(1+& iH_{full}\Delta t/2\hbar) |\psi(t+\Delta t)\rangle\\ &= (1-iH_{full}\Delta t/2\hbar) |\psi(t)\rangle 
+O((H_{full}\Delta t)^3)
\label{crank}
\end{split}
\end{equation}
\noindent
The present time-evolution scheme is also less memory intensive and faster compared to the CN scheme; yet, the 
initial few steps of the evolution is carried out using the CN method. The validity of the above scheme is also 
examined by comparing the time evolution of small systems, calculated by exact methods like either evolving the 
initial state using the matrix representation of $\exp(-iH_{full}\Delta t)$ or by projecting the initial state 
on the eigenstates of $H_{full}$ and explicitly evolving these eigenstates using their corresponding eigenvalues.

The Hamiltonian matrix used for the largest system in our study (16 carbon atoms) is of dimension $\sim 166$ 
million and for reasonable convergence, $\Delta t$ of the order of $0.002\mbox { eV }/\hbar$ is used for the 
PPP model, which is typically $\sim 0.00132$ fs; for H\"uckel and Hubbard model, $\Delta t$ is taken as 
$0.01\mbox { eV }/\hbar$ ~$(\sim 0.0066$ fs). Hence, to follow the dynamics for just $30$ fs, the time evolution 
has to be carried out for more than $20000$ time steps within the PPP model and nearly $5000$ time steps within 
the other two models.

After each time evolution step, the evolved state is projected onto the desired direct product of the triplet 
eigenstates of I and II i.e. $T_m^{I} \otimes T_n^{II}$, where $T_m$ and $T_n$ are triplet eigenstates of 
individual monomers. The total $S_z$ value of the wave packet remains unaltered during time evolution, hence, the 
projection on the triplet channel is carried out in the same $S_z$ space; in this case, both monomers in the 
triplet state can have $S_z=0$ or one of them has $S_z=+1 (-1)$ while the other has $S_z=-1 (+1)$. Triplet 
eigenstates of individual monomers are calculated in $S_z=+1$ space using VB basis and employing $\hat S^-$
operator, corresponding eigenstates in $S_z=0$ and $-1$ spaces are obtained. The yield in a given pair of triplet 
eigenstates $(m,n)$ is given by 
$I_{m,n}(t)=\left| \langle \psi(t)|T_m^{I} \otimes T_n^{II}\rangle_{m,n} \right|^2$ where 
$|T_m^{I} \otimes T_n^{II}\rangle=\frac{1}{\sqrt{3}}|T_{m,S_z=0}^{I} \otimes T_{n,S_z=0}^{II}\rangle 
-\frac{1}{\sqrt{3}}[|T_{m,S_z=+1}^{I} \otimes T_{n,S_z=-1}^{II}\rangle + |T_{m,S_z=-1}^{I} \otimes T_{n,S_z=+1}^{II}\rangle]$, 
according to angular momentum algebra. 
However, the number of such pairs for a neutral subsystem can be enormous; number of triplet-triplet channels for 
octatetraene is $5,531,904$ as each molecule has $2362$ triplet states. Hence, the number of pairs to be 
investigated in each dimer system needs to be significantly reduced. This is achieved by restricting to $\sim 10$ 
low-lying triplet states on each of the neutral subsystems (corresponding to one hundred channels) and by applying
a cut-off in the yield $(\sim 10^{-3})$, which a channel must have at least at one step during the course of the 
full evolution.

\begin{figure*}[p]
\centering
\begin{minipage}[t]{0.48\textwidth}
\centering
\includegraphics[width=\textwidth]{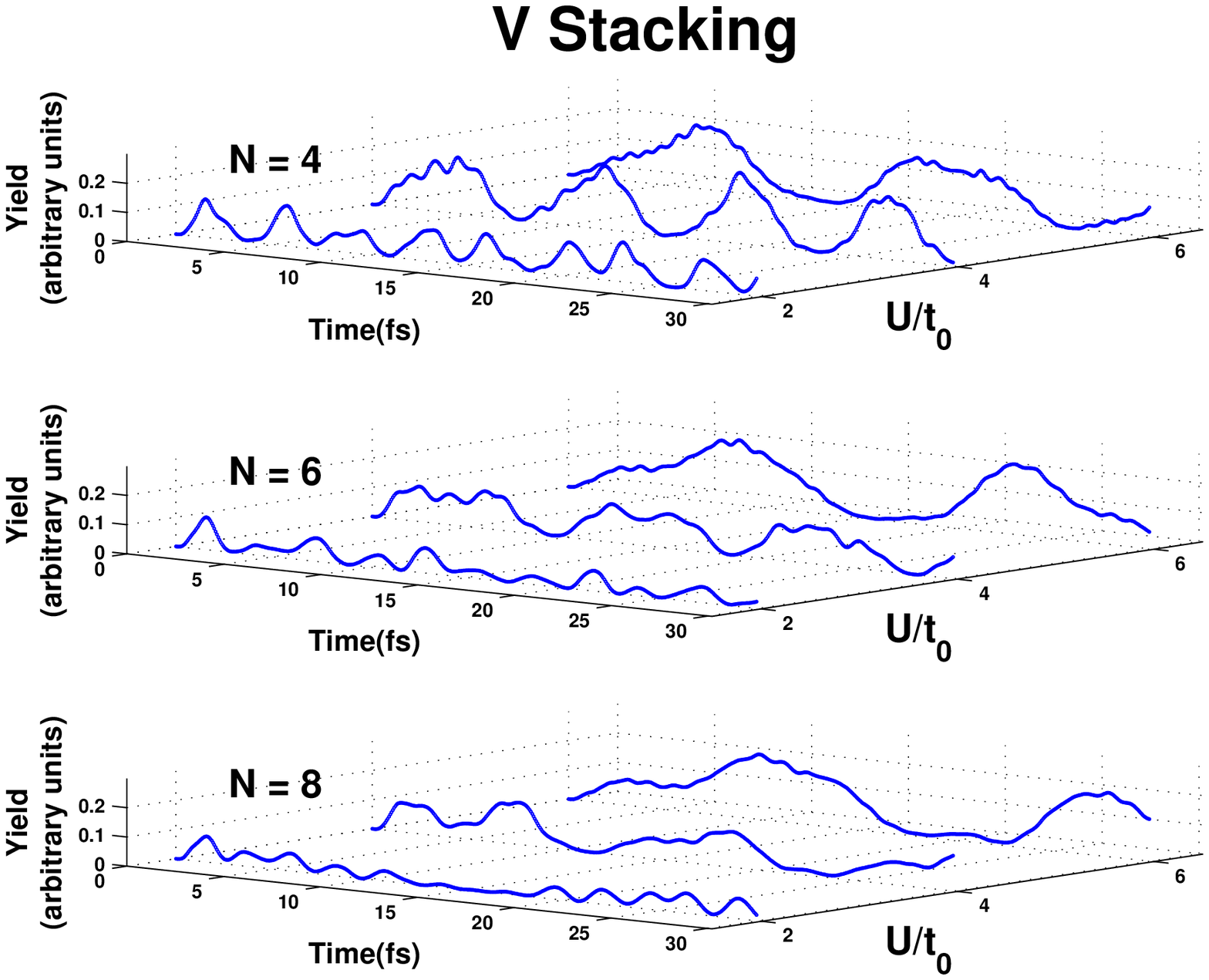}
\LARGE(A)
\end{minipage}
\quad
\begin{minipage}[t]{0.48\textwidth}
\includegraphics[width=\textwidth]{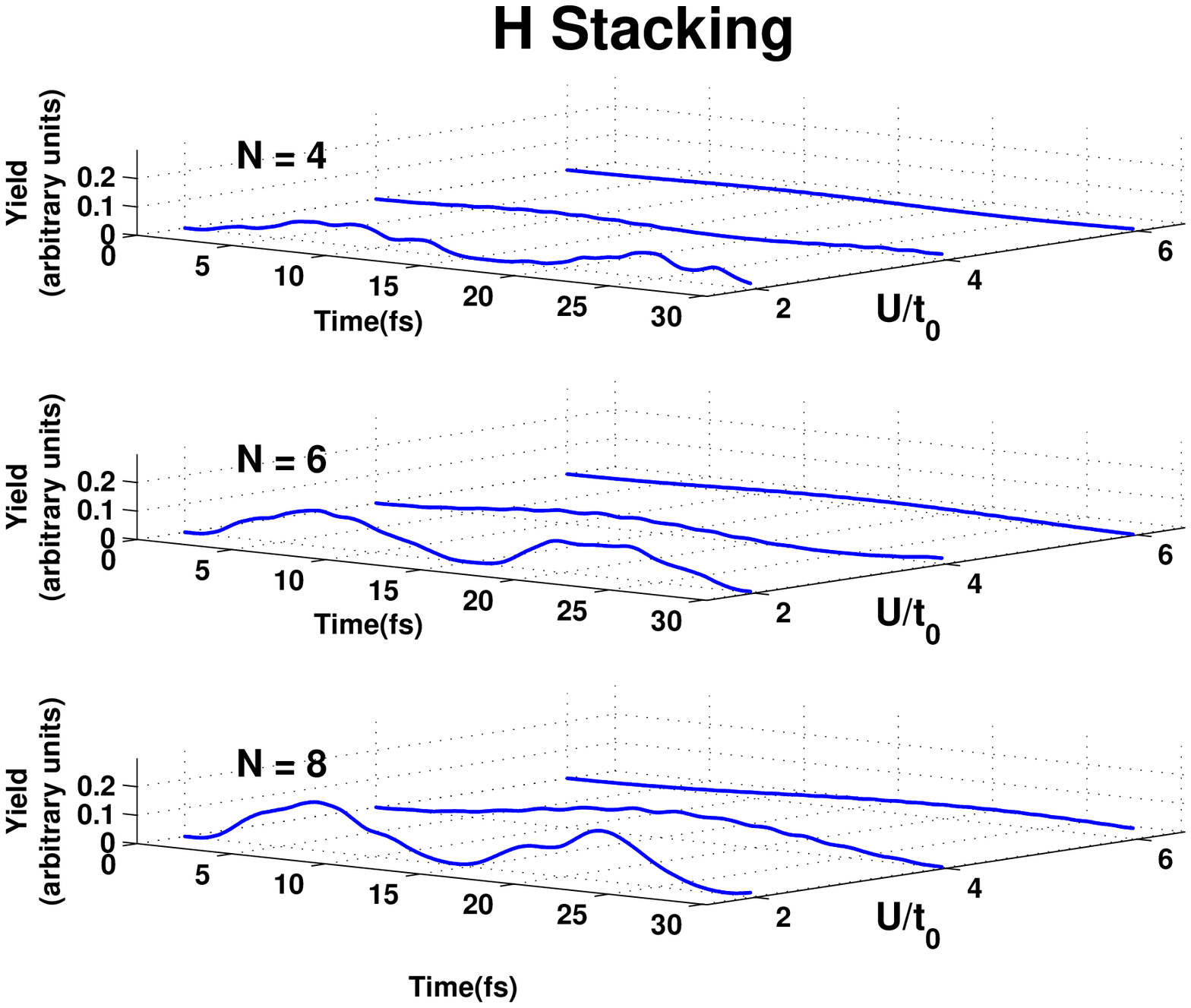}
\LARGE(B)
\end{minipage}
\caption{\footnotesize{(Color online) Time evolution profiles of different polyene dimers in V stacking 
$(t_\perp>0)$ (A) and H stacking $(t_\perp<0)$ (B) for different correlation strengths $(U/t_0)$; here, 
$S_n\equiv 2{}^1A$ case is considered.}} 
\label{evolution_hubb}

\vspace{2.0cm}

\begin{minipage}[t]{0.8\textwidth}
\includegraphics[width=\textwidth,height=0.65\textwidth]{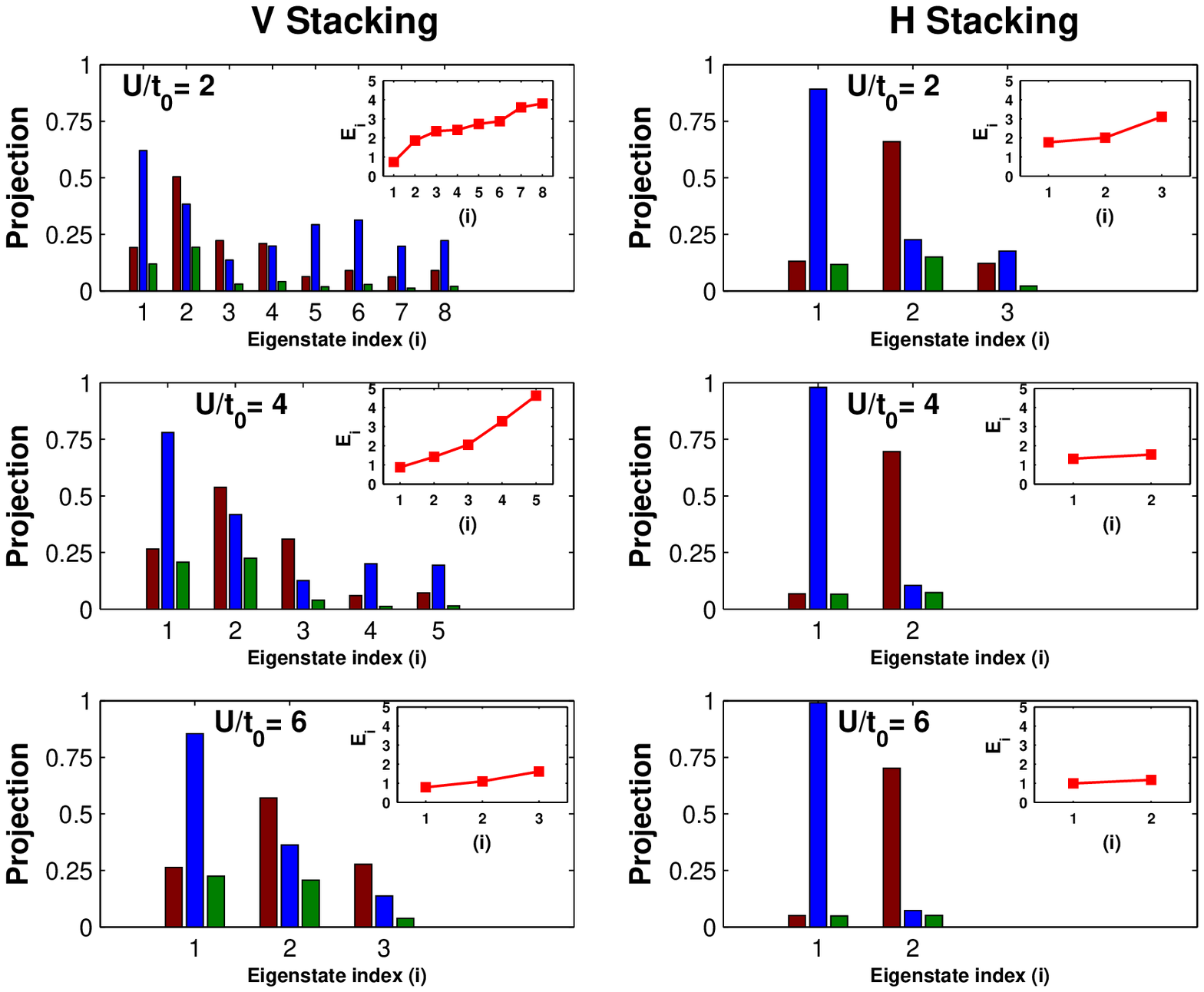}
\caption{\footnotesize{(Color online) Significant projections of $2{}^1A\otimes 1{}^1A$ and $T_1\otimes T_1$ with 
full system eigenstates within Hubbard model are shown as histograms. The left panel 
corresponds to V stacking while the right panel corresponds to H stacking. The color indices are as follows: 
dark brown, projection to initial state, $P_i\equiv \langle 2{}^1A\otimes 1{}^1A|\psi_i\rangle$; 
dark blue, projection to final state, $P_f\equiv \langle T_1\otimes T_1|\psi_i\rangle$; 
dark green, $P_i\times P_f$. Inset: $E_i$, the energy of the significant eigenstate `i' as measured from the 
ground state of the full system is shown.}}
\label{projctn_hubb}
\end{minipage}

\end{figure*}

\section {Results and Discussion}

\begin{table*}[tbp]
\begin{center}
\caption{\label{S_1S_0yield} \footnotesize{Dependence of total yield $(I^{total})$ on the parameters of 
$H_{inter}$ in the Hubbard and PPP model for a pair of butadiene and hexatriene and octatetraene. 
$X_\perp$ is the site-charge--bond-charge repulsion term which is either zero or $0.2$ eV for Hubbard model 
and $0.25$ eV for PPP model. $t_\perp$ is the intermolecular transfer term between corresponding sites 
and within Hubbard model in units of $t_0$, $t_\perp=+0.2$ in V stacking and $t_\perp=-0.2$ in H stacking.
In the PPP model, we have taken $t_\perp=0.25$ eV in V stacking and $t_\perp=-0.25$ eV in H stacking.}}
\begin{ruledtabular}
\renewcommand{\arraystretch}{1.2}
\begin{tabular}{c|ccc|ccc|cc|cc}
System & Model & $U/t_0$ & & \multicolumn{2}{c}{$X_\perp=0$, $t_\perp \ne 0$} & & \multicolumn{2}{c|}{$X_\perp \ne 0$, $t_\perp=0$} & 
\multicolumn{2}{c}{$X_\perp \ne 0$, $t_\perp \ne 0$}\\
\colrule
& & & & V & H & & V & H & V & H \\
\colrule
& & $2.0$ & & $1.70$ & $1.70$ & & $2.32$ & $2.32$ & $1.62$ & $1.18$ \\  
butadiene & Hubbard & $4.0$ & & $0.66$ & $0.66$ & & $2.45$ & $2.45$ & $2.90$ & $0.30$ \\  
& & $6.0$ & & $0.26$ & $0.26$ & & $1.84$ & $1.84$ & $2.65$ & $0.13$ \\ 
\cline{2-11}
& PPP & & & $4.72$ & $4.72$ & & $4.77$ & $4.77$ & $4.42$ & $2.71$ \\ 
\hline
& & $2.0$ & & $2.47$ & $2.47$ & & $2.01$ & $2.01$ & $0.86$ & $1.91$ \\
hexatriene & Hubbard & $4.0$ & & $1.11$ & $1.11$ & & $2.93$ & $2.93$ & $2.12$ & $0.59$ \\
& & $6.0$ & & $0.64$ & $0.64$ & & $2.47$ & $2.47$ & $3.21$ & $0.35$ \\  
\cline{2-11}
& PPP & & & $5.89$ & $5.85$ & & $4.59$ & $4.61$ & $4.67$ & $4.18$ \\
\hline
& & $2.0$ & & $2.98$ & $2.98$ & & $1.32$ & $1.32$ & $0.80$ & $2.34$ \\
octatetraene & Hubbard & $4.0$ & & $2.13$ & $2.13$ & & $3.18$ & $3.18$ & $1.84$ & $1.47$ \\
& & $6.0$ & & $1.57$ & $1.57$ & & $3.10$ & $3.10$ & $3.16$ & $0.89$ \\  
\cline{2-11}
& PPP & & & $8.19$ & $8.23$ & & $5.56$ & $5.57$ & $4.90$ & $6.03$ \\
\end{tabular}
\end{ruledtabular}
\end{center}
\end{table*}

We have computed the integrated yield over the time period of evolution, defined as 
$I_{m,n}^{total}=\sum_iI_{m,n}(t_i)\Delta t$, where $I_{m,n}(t_i)$ is the yield at $i$-th step in triplet pair 
channel $(m,n)$ and $\Delta t$ is the time interval. Our model deals with static nuclei and hence vibronic or 
diabatic effects are ignored. We consider only the primary charge transfer process between two static molecules 
and the product associated with the process; the long-range interacting model is exactly solved with these 
caveats. The initial wave packet is not an eigenstate of the full Hamiltonian and hence evolves with time 
non-trivially under the influence of intermolecular interactions. During time evolution, the total energy of the 
wave packet is not conserved and the wave packet acquires non-zero components of the higher excited states 
through intermolecular interactions; however, the weights of these components are negligible. Therefore, we have 
ignored yields in these unphysical states. We have also observed Rabi type oscillations \cite{Rabi37} expected 
from non-dissipative quantum dynamics. However, physically important final state is the $T_1 \otimes T_1$ state 
and we focus only on this state in all our further discussions. As we have considered only $I_{1,1}(t)$, the 
subscript is dropped in all later discussions.
 

In our study, we have considered two different choices of the $S_n$ state.
In the first, we have considered $S_n\equiv S_1$; i.e., the lowest energy singlet excited state of one monomer
and ground state of the other monomer is employed in constructing the initial wave packet. 
On the other hand, in the second, the lowest optical state is considered as the $S_n$ state 
$(S_n\equiv S_{op})$. Substitution by donor-acceptor groups at the end of the chains breaks spatial symmetry 
$(C_2)$ as well as electron-hole symmetry and results in mixing of eigenstates of different symmetries of the 
unsubstituted system; consequently, every eigenstate becomes optically allowed on substitution. In this case, 
we have considered the state with highest transition dipole moment from the ground state (within an energy window) 
as $S_{op}$, i.e $S_n\equiv S_{op}=S_{\mu_{tr}^{max}}$. In all substituted polyenes, it has been assumed that 
the donor and acceptor strengths are same, i.e. $|\epsilon_D|=|\epsilon_A|=\epsilon$. For large enough 
$\epsilon$, $S_1$ and $S_{op}$ states become same and there remains no difference between the two scenarios.  

According to the Fermi's golden rule, the rate of transition probability from the initial state $|i\rangle$ to 
the final state $|f\rangle$ is given by:

\begin{equation}
W_{i\rightarrow f}=\frac{2\pi}{\hbar}|\langle i|H_{inter}|f\rangle|^2 \rho_f(E_f)
\label{fermi_pic}
\end{equation}
\noindent
In the case of polyenes the density of excitonic states is given by $\delta(E-E_f)$ as the spectrum is discrete.
This implies that the transition rate is completely governed by the matrix element $\langle i|H_{inter}|f\rangle$.
We have computed the matrix element in both non-interacting and interacting models and found it to be negligible. 
Thus, within this simple approach, we will not observe any SF. 

To study the SF process in detail and obtain physical insights, we have analyzed small polyene systems (dimer of 
unsubstituted or substituted 1,3-butadiene). We express the initial wave packet $\Psi(0)$ as a linear combination 
of the eigenstates $\psi_k$ of the full Hamiltonian with eigenvalues $E_k$. The time evolution of the wavepacket 
is carried out using the eigenvalues of the corresponding states, 
i.e., $\Psi(t)=\sum_k c_k|\psi_k\rangle \exp(-iE_kt/\hbar)$. The yield $I(t)$ in this approach is given by:

\begin{equation}
\begin{split}
I(t) &= \sum_i |\langle S_n\otimes S_0|\psi_i(0)\rangle \langle T_1\otimes T_1|
\psi_i(0)\rangle|^2 \\ &+ 2\sum_i\sum_{j>i} \mbox{Re} \left\{ \langle S_n\otimes S_0|\psi_i(0)\rangle \langle T_1\otimes T_1|
\psi_i(0)\rangle \right.\\&\left. \langle \psi_j(0)|S_n\otimes S_0\rangle \langle \psi_j(0)|T_1\otimes T_1\rangle \right\} \cos(\omega_{ij}t)
\end{split}
\label{exact_evoln}
\end{equation}
\begin{equation*}
\omega_{ij}\equiv (E_i-E_j)/\hbar
\end{equation*}

\noindent
From Eq. \ref{exact_evoln}, it can be noted that for a high cross-section in singlet fission, it is necessary 
that at least one of the eigenstates of the full system Hamiltonian should have simultaneously large non-zero
overlaps with the initial and final states. 

In the H\"uckel model, the lowest energy excited state is also the optical state $(S_1\equiv1{}^1B)$. The 
two-photon state remains much higher in energy compared to the optical state and we do not consider evolution 
from this state. The energies of the lowest optical state and the lowest triplet state are same in the H\"uckel 
picture. Thus energetically a single optically excited molecule cannot yield two triplets in the H\"uckel model. 
Indeed, yields in the $T_1\otimes T_1$ channel are zero for both V and H stackings (Fig. \ref{system}). The 
overlap integrals of the full system eigenstates with the initial and final states are not large simultaneously 
(Fig. $S1$ in the supplemental material \cite{supple}), which is a prerequisite for large yield. Hence, without electron 
correlations, the cross-section for singlet fission will be negligible.

We have studied singlet fission for a pair of butadienes, hexatrienes and octatetraenes within the Hubbard model 
for different on-site correlation strengths, $U/t_0$. In order to understand the role of $X_\perp$ term vis a vis 
that of $t_\perp$, we have studied three cases (i) $X_\perp=0; t_\perp \ne 0$, (ii) $X_\perp \ne 0; t_\perp=0$ and
(iii) $X_\perp \ne 0; t_\perp \ne 0$ with $2{}^1A$ state as the initial singlet excited state 
(Table. \ref{S_1S_0yield}). In cases (i) and (ii), we find that the yield does not depend upon the type of 
stacking. However, the $X_\perp$ term gives rise to higher SF yield compared to $t_\perp$ term at larger $U/t_0$ 
values. In case (iii), when both $X_\perp$ and $t_\perp$ are nonzero, we find a synergistic effect on the SF 
yield in both stacking orientations. In the V stacking, the SF yield increases with correlation strength while 
in the H stacking it decreases with correlation strength. In the case of PPP model, we find that the 
yields are significantly larger than in the Hubbard model. Furthermore, the yield increases with chain length, 
showing the importance of intermolecular interactions. In the case of hexatriene and octatetraene dimers, 
$X_\perp$ term leads to a decrease in the yield in all cases. When both $X_\perp$ and $t_\perp$ are present, the 
yield is marginally higher for H stacking than in V stacking for longer oligomer.

To understand this behavior, we have focused on the bond order $(-\langle E_{ii'}+ E_{i'i} \rangle /2)$ between 
corresponding sites of the two molecules. The bond order is larger when $t_\perp<0$ and smaller when $t_\perp>0$. 
The larger bond order implies the site-charge densities are more uniform in the eigenstates of the full system. 
This leads to smaller contribution from the $X_\perp$ term in H stacking since the amplitude for hopping 
due to $X_\perp$ term is site-charge dependent. 

\begin{figure}[b]
\centering
\includegraphics[height=7.5cm]{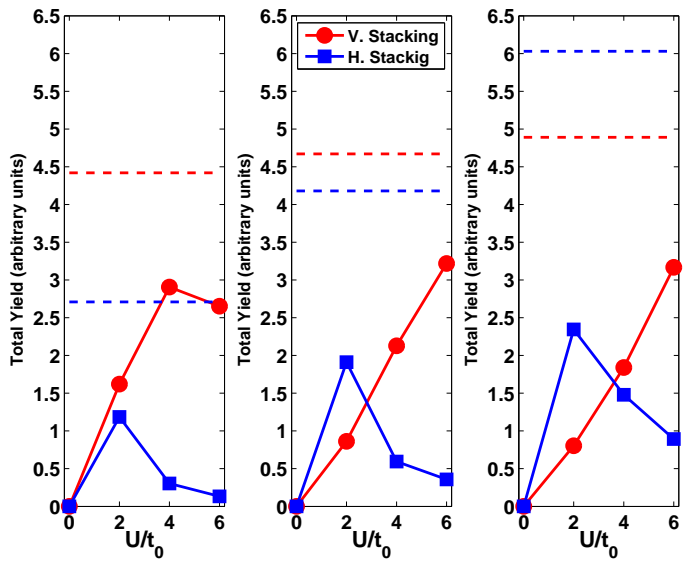}
\caption{\footnotesize{(Color online)$I^{total}$ is plotted as a function of correlation strength $U/t_0$ for 
$S_n\equiv 2{}^1A$. The left, center and right panels correspond to monomers of $4$, $6$ and $8$ sites. 
Red filled circle corresponds to V stacking and blue filled square corresponds to H stacking.
The broken lines in each panel correspond to the PPP values for unsubstituted system in V (red) and H (blue) 
stackings. The solid lines are given only as a guide to the eye.}}
\label{totlyld_hubb}
\end{figure}

When $S_n\equiv1{}^1B$, total yields are insignificant in both V and H stackings. Analysis employing full system 
Hamiltonian eigenstates shows highly disjoint overlaps with the initial and final states (Fig. S2 in the 
supplemental material \cite{supple}), similar to the H\"uckel model. Hence, only choice of $2{}^1A$ for the initial excited 
singlet state results in significant $I^{total}$ in both stackings (Fig. \ref{evolution_hubb} and 
\ref{totlyld_hubb}).

The time evolution profiles, shown in Fig. \ref{evolution_hubb}(A) and \ref{evolution_hubb}(B) shed light on the 
dependence of $I^{total}$ on monomer chain length in the Hubbard model. In V stacking, at a particular $U/t_0$, 
temporal variation of $I(t)$ in the evolution profile becomes weaker for longer chain systems, and the 
oscillatory pattern becomes more complex. In these cases, the eigenspectrum of dimers become more dense with 
increasing chain length and larger number of eigenstates contribute significantly towards $I(t)$ resulting in 
complex interference in the time evolution profile. The time evolution profiles (Fig. \ref{evolution_hubb}(B)) 
also suggest that the significant eigenstates in H stacking are almost degenerate as the yield shows simpler 
time dependence. This can also be seen from the right panel in Fig. \ref{projctn_hubb}.

\begin{table*}[tbp]
\begin{center}
\caption{\label{projctn_e0} \footnotesize{Full system eigenstates of 1,3-butadiene dimer having significant 
projections with $S_n\otimes 1{}^1A$ $(P_i)$ and $T_1\otimes T_1$ $(P_f)$ in the PPP model are tabulated for 
V and H stacking. $E$ is the excitation energy of the full system in eV.}}
\begin{ruledtabular}
\renewcommand{\arraystretch}{1.2}
\begin{tabular}{cccc|cccc|cccc|cccc}
\multicolumn{8}{c|}{V stacking} & \multicolumn{8}{c}{H stacking}  \\
\cline{1-16}
\multicolumn{4}{c|}{$S_n\equiv2{}^1A$} & \multicolumn{4}{c|}{$S_n\equiv1{}^1B$} & \multicolumn{4}{c|}{$S_n\equiv2{}^1A$} & \multicolumn{4}{c}{$S_n\equiv1{}^1B$} \\
\cline{1-16}
$E$ & $P_i$ & $P_f$ & $P_i\times P_f$ & $E$ & $P_i$ & $P_f$ & $P_i\times P_f$ & 
$E$ & $P_i$ & $P_f$ & $P_i\times P_f$ & $E$ & $P_i$ & $P_f$ & $P_i\times P_f$  \\
\colrule
$4.71$ & $0.3$ & $0.8$ & $0.2$ & $4.71$ & $0.0$ & $0.8$ & $0.0$ & 
$5.28$ & $0.2$ & $0.9$ & $0.2$ & $5.28$ & $0.0$ & $0.9$ & $0.0$ \\

$5.32$ & $0.6$ & $0.5$ & $0.3$ & $4.93$ & $0.6$ & $0.0$ & $0.0$ & 
$5.34$ & $0.6$ & $0.4$ & $0.2$ & $5.34$ & $0.0$ & $0.4$ & $0.0$ \\

       &       &       &       & $5.31$ & $0.0$ & $0.5$ & $0.0$ &
       &       &       &       & $5.56$ & $0.7$ & $0.0$ & $0.0$ \\

       &       &       &       & $5.84$ & $0.7$ & $0.0$ & $0.0$ &
       &       &       &       & $5.90$ & $0.7$ & $0.0$ & $0.0$ \\

       &        &        &        & $7.55$ & $0.2$ & $0.0$ & $0.0$ &
        &        &        &        &       &       &       &       \\

       &        &        &        & $8.18$ & $0.3$ & $0.0$ & $0.0$ &
        &        &        &        &       &      &      &      \\

       &        &        &        & $8.73$ &$0.0$& $0.2$ &$0.0$&
        &        &        &        &        &        &        &        \\

       &        &        &        & $11.04$ & $0.0$ & $0.1$ & $0.0$ &
        &        &        &        &        &        &        &        \\

\end{tabular}
\end{ruledtabular}
\end{center}
\end{table*}

\begin{figure}[bh]
\begin{center}
\includegraphics[height=6.5cm]{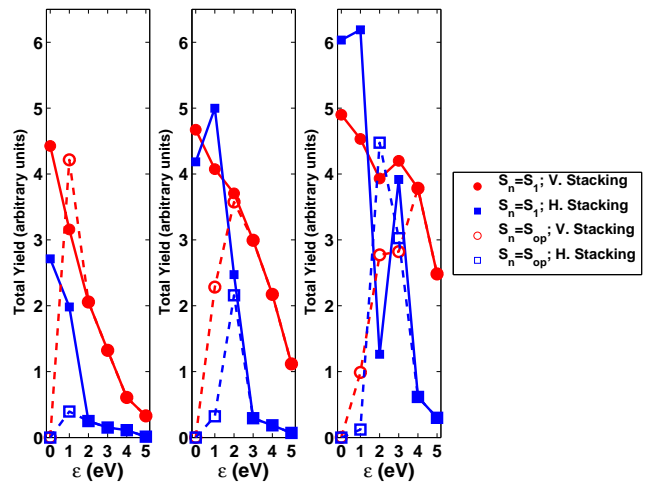}
\caption{\footnotesize{(Color online) Total yield in unsubstituted and substituted polyene dimers are plotted as 
a function of substitution strength $\epsilon$ for both vertical and horizontal stacking orientations within the 
PPP model. The left, center and right panels correspond to monomers of $4$, $6$ and $8$ sites. The color and 
symbol indices are given in the following and same in all three panels: red filled circle, $S_n\equiv S_1$, 
V stacking; blue filled square, $S_n\equiv S_1$, H stacking. The open symbols represent $S_n\equiv S_{op}$ 
scenarios in the corresponding systems. Beyond a certain $\epsilon$, $S_n\equiv S_1\equiv S_{op}$ and the curves 
coincide. The solid and broken lines are given only as a guide to the eyes.}}
\label{totlyld}
\end{center}
\end{figure}

Organic systems that we are interested in are semiconducting. Hence long-range interactions are not screened out 
as in metals and for a realistic modeling of the system we need to include explicit long-range electron-electron 
interactions. The PPP model with standard parameters is well suited for modeling conjugated organics 
{\cite{schulten79-1,*schulten79-2,*schulten86,Schulten87,Soos84,Soos93,Baeriswyl}}. We have found that 
introducing long-range interaction dramatically changes the yield of triplets. When the wave packet is built 
from an optical state $(1{}^1B)$ on one molecule and ground state on another, the total yield remains quite low 
in both V and H stackings. On the other hand, when the initial wave packet is constructed from $2{}^1A$ state 
and the ground state, there is significant increase in $I^{total}$, as can be seen from Figs. \ref{totlyld_hubb} 
and \ref{totlyld}. For polyenes within the PPP model, large number of eigenstates have significant simultaneous 
projections on both $|2{}^1A\otimes 1{}^1A\rangle$ and $|T_1\otimes T_1\rangle$ (Table. \ref{projctn_e0}); these 
states are also nearly isoenergetic in H stacking leading to constructive interference (Eq. \ref{exact_evoln}) 
and large yields (Fig. \ref{evolution_16} and Figs. S3 and S4 in the supplemental material \cite{supple}). In V stacking, on 
the other hand, the contributing states have different energies and the yield is lower. 
It should be noted that this conclusion excludes effects of molecular vibrations or phonons on the SF process.

\begin{figure*}[tp]
\centering
\begin{minipage}[t]{0.48\textwidth}
\centering
\includegraphics[width=\textwidth,height=1.0\textwidth]{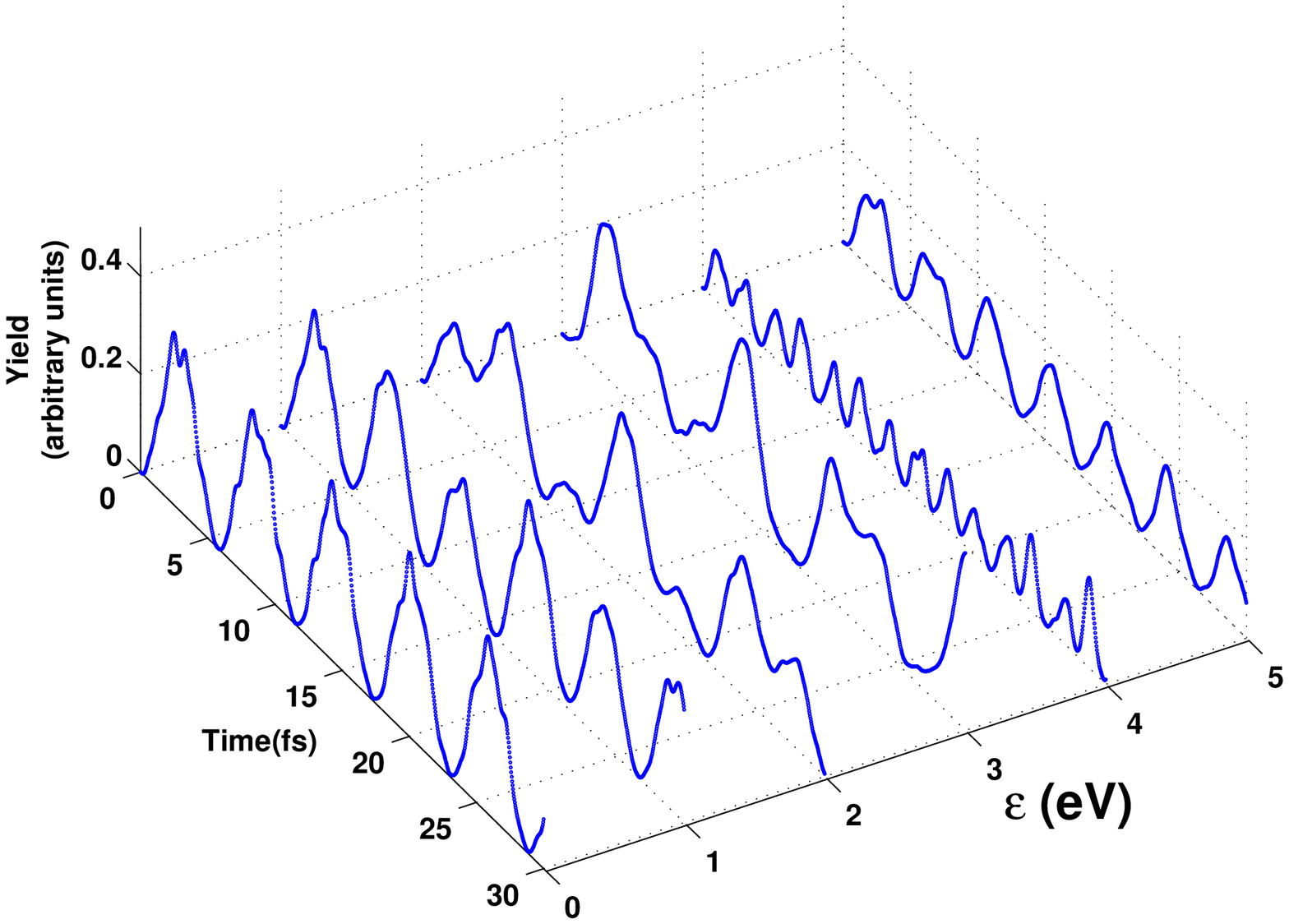}
\Large(A)
\end{minipage}
\quad
\begin{minipage}[t]{0.48\textwidth}
\centering
\includegraphics[width=\textwidth,height=1.0\textwidth]{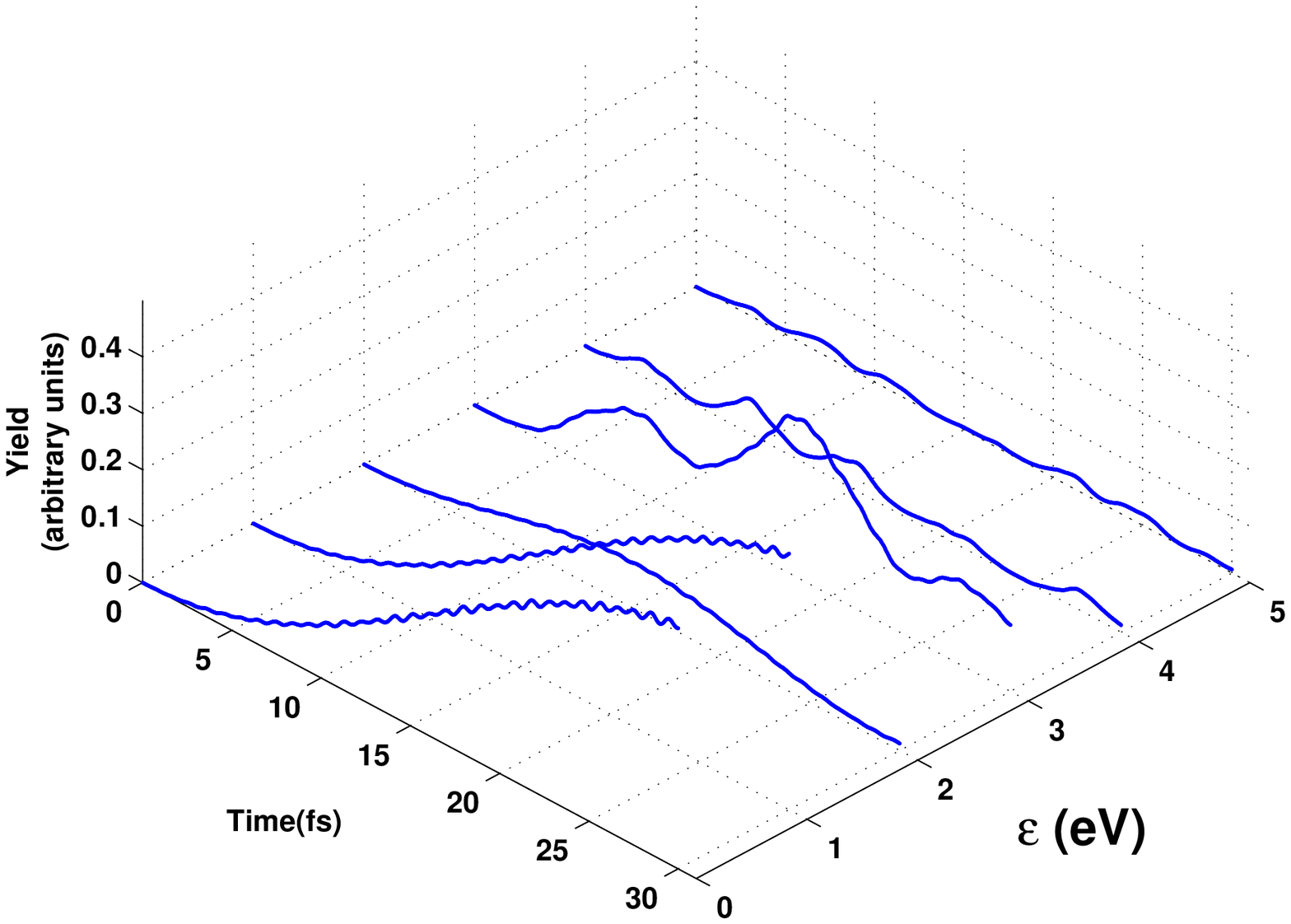}
\Large(B)
\end{minipage}

\vspace{1.5cm}

\begin{minipage}[t]{0.48\textwidth}
\centering
\includegraphics[width=\textwidth,height=1.0\textwidth]{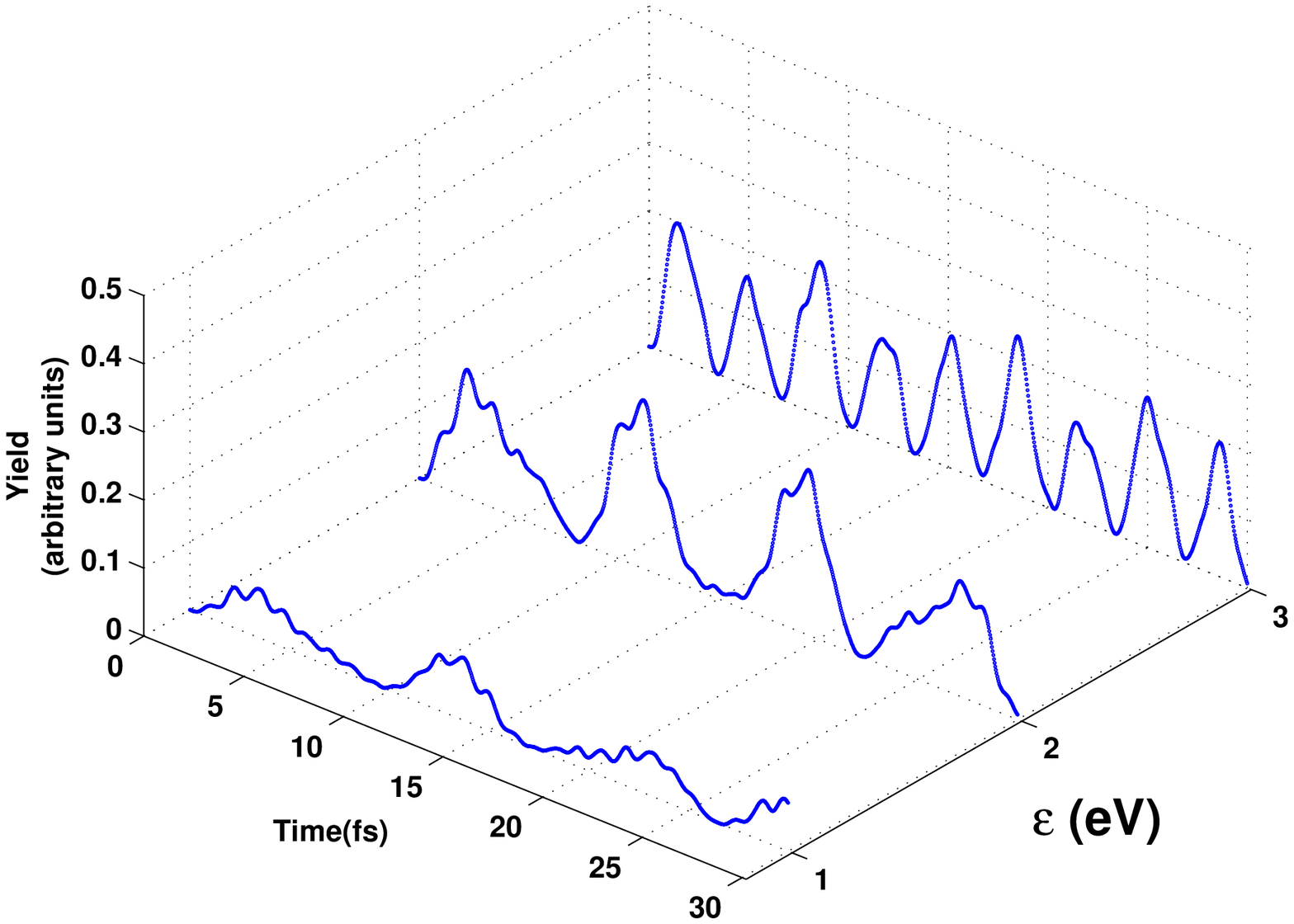}
\Large(C)
\end{minipage}
\quad
\begin{minipage}[t]{0.48\textwidth}
\centering
\includegraphics[width=\textwidth,height=1.0\textwidth]{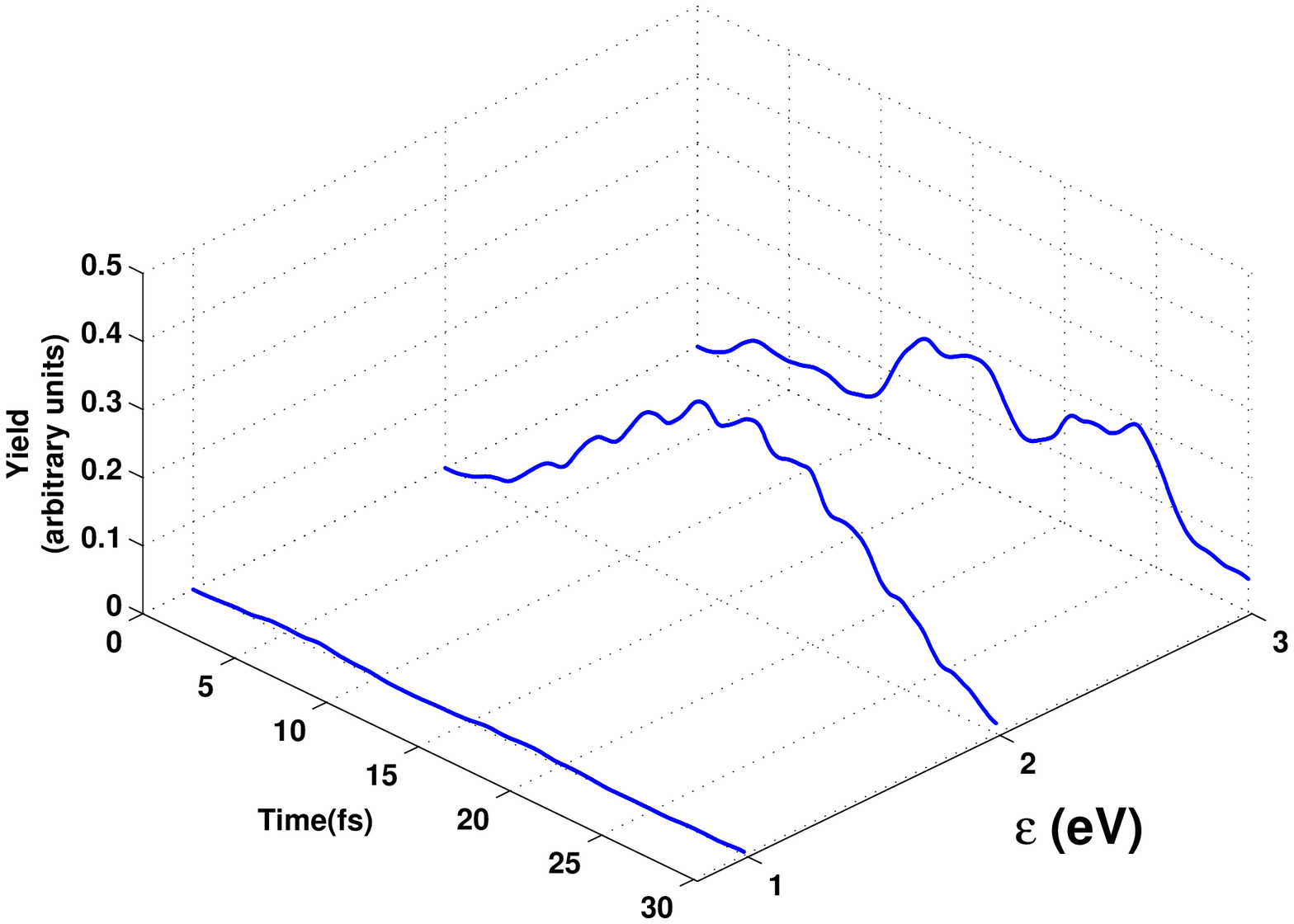}
\Large(D)
\end{minipage}
\begin{minipage}[c]{0.48\textwidth}
\vspace{3.0cm}
\end{minipage}
\caption{\footnotesize{(Color online) Yield as a function of time and donor-acceptor strength $\epsilon$ for 
singlet fission in 1,3,5,7-octatetraene dimer from the lowest singlet excited state $S_1$ in (A) V stacking and 
(B) H stacking. For $\epsilon=1$ eV, $2$ eV and $3$ eV, the yield from the optical singlet state is also shown 
in (C) V stacking and (D) H stacking. For $\epsilon \ge 4$ eV, we find that the lowest excited state is also the 
state with large transition dipole moment.}}
\label{evolution_16}
\end{figure*}

\begin{table*}[tb]
\begin{center}
\caption{\label{energap} \footnotesize{Energy gaps in butadiene, hexatriene and octatetraene within H\"{u}ckel, 
Hubbard and PPP models. Energy gaps are given in units of $t_0$ within H\"uckel and Hubbard models.}}
\begin{ruledtabular}
\renewcommand{\arraystretch}{1.2}
\begin{tabular}{ccddddd}
& & \multicolumn{1}{c}{$U/t_0=0$} & \multicolumn{1}{c}{$U/t_0=2$} & \multicolumn{1}{c}{$U/t_0=4$} 
& \multicolumn{1}{c}{$U/t_0=6$} & \multicolumn{1}{c}{PPP (eV)} \\
\colrule
& $E_{1{}^1B}$ & $1.40$ & $2.19$ & $3.40$ & $4.94$ & $5.83$ \\
& $E_{2{}^1A}$ & $2.33$ & $2.06$ & $1.54$ & $1.17$ & $5.34$ \\
butadiene & $E_{T_1}$ & $1.40$ & $0.96$ & $0.67$ & $0.50$ & $2.67$ \\
& $E_{1{}^1B}-2E_{T_1}$ & $-1.40$ & $0.27$ & $2.06$ & $3.94$ & $0.49$ \\
& $E_{2{}^1A}-2E_{T_1}$ & $-0.47$ & $0.14$ & $0.20$ & $0.17$ & $0.00$ \\
\colrule
& $E_{1{}^1B}$ & $1.07$ & $1.68$ & $2.78$ & $4.26$ & $5.05$ \\
& $E_{2{}^1A}$ & $1.81$ & $1.63$ & $1.25$ & $0.96$ & $4.36$ \\
hexatriene & $E_{T_1}$ & $1.07$ & $0.76$ & $0.56$ & $0.42$ & $2.18$ \\
& $E_{1{}^1B}-2E_{T_1}$ & $-1.07$ & $0.16$ & $1.66$ & $3.42$ & $0.69$ \\
& $E_{2{}^1A}-2E_{T_1}$ & $-0.33$ & $0.11$ & $0.13$ & $0.12$ & $0.00$ \\
\colrule
& $E_{1{}^1B}$ & $0.88$ & $1.39$ & $2.44$ & $3.90$ & $4.56$ \\
& $E_{2{}^1A}$ & $1.47$ & $1.37$ & $1.07$ & $0.83$ & $3.75$ \\
octatetraene & $E_{T_1}$ & $0.88$ & $0.66$ & $0.49$ & $0.38$ & $1.90$ \\
& $E_{1{}^1B}-2E_{T_1}$ & $-0.88$ & $0.07$ & $1.46$ & $3.14$ & $0.76$ \\
& $E_{2{}^1A}-2E_{T_1}$ & $-0.29$ & $0.05$ & $0.09$ & $0.07$ & $-0.05$ \\
\end{tabular}
\end{ruledtabular}
\end{center}
\end{table*}

The total yield for varying $\epsilon$ is plotted in Fig. \ref{totlyld} while the time evolution profiles for 
octatetraene dimers for different substitution strength are shown in Fig. \ref{evolution_16}; corresponding time 
evolution profiles for butadiene and hexatriene are given in Figs. S3 and S4 in the supplemental material \cite{supple}. 
In V stacking, $I^{total}$ decreases with increasing substitution strength when $S_n\equiv S_1$ and this outcome is 
independent of monomer size (except for octatetraene dimer with $\epsilon=3.0$). 
Yet, when $S_{op}$, the state to which the transition dipole moment is largest is considered, 
non-monotonous behavior of $I^{total}$ is observed with increasing $\epsilon$. The total yield is dependent on 
the nature of the singlet excited state at small $\epsilon$. At large $\epsilon$ the lowest excited singlet state 
is also the most strongly optically allowed singlet state and the distinction ceases. In contrast, H stacking 
orientation exhibits unique $I^{total}$ profile with increasing $\epsilon$ (Fig. \ref{totlyld}); large variation 
in $I^{total}$ as a function of $\epsilon$ is observed for both $S_n\equiv S_1$ and $S_n\equiv S_{op}$ cases. 
This general trend is observable in all three polyene systems considered.

\begin{figure}[h]
\centering
\includegraphics[height=5.6cm,width=7.8cm]{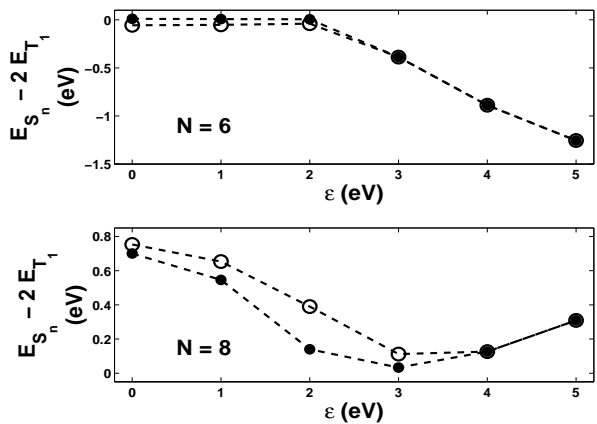}
\caption{\footnotesize{Energy difference between the initial and final coupled states $(E_{S_n}-2E_{T_1})$ within 
PPP model are plotted against various $\epsilon$ for hexatriene and octatetraene. Filled circles correspond to 
$S_N\equiv S_1$ while open circles correspond to $S_n\equiv S_{op}$ scenario. For $\epsilon=0$, 
$S_1\equiv 2{}^1A$ while $S_{op}$ is the lowest energy state of $B$ symmetry subspace $(S_{op}\equiv 1{}^1B)$. 
In substituted PPP dimers, beyond a particular $\epsilon$, $S_1$ becomes equivalent with $S_{op}$ and the curves 
coincide. The broken lines are shown only as a guide to the eyes.}}
\label{energap_ppp}
\end{figure}

The energetics show that the singlet fission process is either slightly exoergic or endoergic for substituted 
PPP chains (Fig. \ref{energap_ppp}). From energy consideration, we note that donor-acceptor strength $\epsilon$
between $2.0$ and $3.0$ eV would result in large  SF yield.

\section{Conclusion}

The energy criteria proposed by Michl et al. (Ref. \citenum{Michl10,Michl13-1,Michl13-2}) that singlet fission is 
feasible when the initial state energy is greater than or equal to the final state energy is seen to be operative 
in our model studies. In the H\"{u}ckel model the optical $1{}^1B$ state is degenerate with the triplet state and 
as a consequence the energy criteria is not met, resulting in insignificant singlet fission yield. The energy of 
the $2{}^1A$ state in the H\"{u}ckel model is also not close to the total energy of two triplets and hence 
$2{}^1A$ will also not yield singlet fission products in non-interacting models.

When electron correlations are turned on, as in the Hubbard or PPP models, the energy of the triplet state, being 
covalent, comes down while the energy of $1{}^1B$ state, being ionic, increases. There is a crossover in the 
$1{}^1B$ and $2{}^1A$ states, depending upon the correlation strength and chain length \cite{Soos93,Shuai97}. The 
energy of the $2{}^1A$ state is nearly twice the energy of lowest triplet state (Table. \ref{energap}) and hence, 
an initial singlet excitation in $2{}^1A$ state yields significant triplets in the singlet fission process. 
However, the $1{}^1B$ state is not energetically close to two triplets and yields insignificant SF products. 
Analysis on butadiene dimer shows that in all models, the simultaneous overlap of the wave packet and the final 
product state with the eigenstates of $H_{full}$ is negligible when the wave packet is formed from $1{}^1B$ state. 
On the other hand, for wave packet constructed from $2{}^1A$ state, the simultaneous overlap is significant for 
some eigenstates, thereby leading to fission products. Indeed, we also find from the analysis of the 
full system eigenstates that there is a singlet excited state of the full system which is a coherent state of the 
$2{}^1A$ singlet and two triplets. Since in reality, the excitations occur in the full system, we can conclude 
that the primary excitation is to an optically allowed excitation, which leads to this coherent state through 
internal conversion. Our studies also show that excitation to $1{}^1B$ state does not directly yield SF products 
as suggested by Musser et al. {\cite{Musser13,Musser15}}. This is also because the $1{}^1B$ excitation is to an 
ionic state while the $2{}^1A$ and the triplets involved in SF are covalent states. Our studies are also in 
agreement with earlier PPP studies which postulate a coherent state {\cite{Aryanpour15-2}} as well as those that 
show the importance of the $2{}^1A$ state {\cite{Shuai17}}.

In substituted polyene chains, for small donor-acceptor strengths singlet state derived from $1{}^1B$ state also 
gives significant fission yield within PPP model due to mixing of $2{}^1A$ state in the eigenstates. For higher 
donor-acceptor strengths, the singlet state derived from the $2{}^1A$ state loses its two-triplet character and 
the fission yield goes down significantly. We have also found that fission yield depends on stacking geometry. In 
V stacking where the intermolecular transfer integral $t_\perp>0$, the singlet fission yield in the PPP model 
decreases with increasing chain length while when $t_\perp<0$ as in H stacking, there is an increase in fission 
yield. We expect SF to occur from the lowest excited singlet state, as fast internal conversions lead to this 
state independent of the initial excited state reached by photoexcitation. Hence, in systems where the lowest 
singlet excited state is the state with large $2{}^1A$ character, we expect significant SF yield. However, if the 
lowest excited singlet state has largely $1{}^1B$ character, the SF yield will be negligible. Thus, we can see 
that systems which are fluorescent will not give large SF yields. Therefore, systems which are good for light 
emission will not be good candidates for improving photovoltaic efficiency through SF.

\begin{acknowledgments}

S.R. is thankful to the Department of Science and Technology, India for financial support through various grants. 
S.P. acknowledges CSIR India for a senior research fellowship. S.P. also thanks Prof. Diptiman Sen for financial 
support through his J. C. Bose fellowship.
\vspace{-0.3cm}

\end{acknowledgments}

\bibliography{manuscript}

\begin{thebibliography}{91}%
\makeatletter
\providecommand \@ifxundefined [1]{%
 \@ifx{#1\undefined}
}%
\providecommand \@ifnum [1]{%
 \ifnum #1\expandafter \@firstoftwo
 \else \expandafter \@secondoftwo
 \fi
}%
\providecommand \@ifx [1]{%
 \ifx #1\expandafter \@firstoftwo
 \else \expandafter \@secondoftwo
 \fi
}%
\providecommand \natexlab [1]{#1}%
\providecommand \enquote  [1]{``#1''}%
\providecommand \bibnamefont  [1]{#1}%
\providecommand \bibfnamefont [1]{#1}%
\providecommand \citenamefont [1]{#1}%
\providecommand \href@noop [0]{\@secondoftwo}%
\providecommand \href [0]{\begingroup \@sanitize@url \@href}%
\providecommand \@href[1]{\@@startlink{#1}\@@href}%
\providecommand \@@href[1]{\endgroup#1\@@endlink}%
\providecommand \@sanitize@url [0]{\catcode `\\12\catcode `\$12\catcode
  `\&12\catcode `\#12\catcode `\^12\catcode `\_12\catcode `\%12\relax}%
\providecommand \@@startlink[1]{}%
\providecommand \@@endlink[0]{}%
\providecommand \url  [0]{\begingroup\@sanitize@url \@url }%
\providecommand \@url [1]{\endgroup\@href {#1}{\urlprefix }}%
\providecommand \urlprefix  [0]{URL }%
\providecommand \Eprint [0]{\href }%
\providecommand \doibase [0]{http://dx.doi.org/}%
\providecommand \selectlanguage [0]{\@gobble}%
\providecommand \bibinfo  [0]{\@secondoftwo}%
\providecommand \bibfield  [0]{\@secondoftwo}%
\providecommand \translation [1]{[#1]}%
\providecommand \BibitemOpen [0]{}%
\providecommand \bibitemStop [0]{}%
\providecommand \bibitemNoStop [0]{.\EOS\space}%
\providecommand \EOS [0]{\spacefactor3000\relax}%
\providecommand \BibitemShut  [1]{\csname bibitem#1\endcsname}%
\let\auto@bib@innerbib\@empty
\bibitem [{\citenamefont {Singh}\ \emph {et~al.}(1965)\citenamefont {Singh},
  \citenamefont {Jones}, \citenamefont {Siebrand}, \citenamefont {Stoicheff},\
  and\ \citenamefont {Schneider}}]{Schneider65}%
  \BibitemOpen
  \bibfield  {author} {\bibinfo {author} {\bibfnamefont {S.}~\bibnamefont
  {Singh}}, \bibinfo {author} {\bibfnamefont {W.~J.}\ \bibnamefont {Jones}},
  \bibinfo {author} {\bibfnamefont {W.}~\bibnamefont {Siebrand}}, \bibinfo
  {author} {\bibfnamefont {B.~P.}\ \bibnamefont {Stoicheff}}, \ and\ \bibinfo
  {author} {\bibfnamefont {W.~G.}\ \bibnamefont {Schneider}},\ }\href@noop {}
  {\bibfield  {journal} {\bibinfo  {journal} {J. Chem. Phys.}\ }\textbf
  {\bibinfo {volume} {42}},\ \bibinfo {pages} {330} (\bibinfo {year}
  {1965})}\BibitemShut {NoStop}%
\bibitem [{\citenamefont {Swenberg}\ and\ \citenamefont
  {Stacy}(1968)}]{Swenberg68}%
  \BibitemOpen
  \bibfield  {author} {\bibinfo {author} {\bibfnamefont {C.~E.}\ \bibnamefont
  {Swenberg}}\ and\ \bibinfo {author} {\bibfnamefont {W.~T.}\ \bibnamefont
  {Stacy}},\ }\href@noop {} {\bibfield  {journal} {\bibinfo  {journal} {Chem.
  Phys. Lett.}\ }\textbf {\bibinfo {volume} {2}},\ \bibinfo {pages} {327}
  (\bibinfo {year} {1968})}\BibitemShut {NoStop}%
\bibitem [{\citenamefont {Merrifield}\ \emph {et~al.}(1969)\citenamefont
  {Merrifield}, \citenamefont {Avakian},\ and\ \citenamefont
  {Groff}}]{Merrifield69}%
  \BibitemOpen
  \bibfield  {author} {\bibinfo {author} {\bibfnamefont {R.~E.}\ \bibnamefont
  {Merrifield}}, \bibinfo {author} {\bibfnamefont {P.}~\bibnamefont {Avakian}},
  \ and\ \bibinfo {author} {\bibfnamefont {R.~P.}\ \bibnamefont {Groff}},\
  }\href@noop {} {\bibfield  {journal} {\bibinfo  {journal} {Chem. Phys.
  Lett.}\ }\textbf {\bibinfo {volume} {3}},\ \bibinfo {pages} {386} (\bibinfo
  {year} {1969})}\BibitemShut {NoStop}%
\bibitem [{\citenamefont {Groff}\ \emph {et~al.}(1970)\citenamefont {Groff},
  \citenamefont {Avakian},\ and\ \citenamefont {Merrifield}}]{Merrifield70-1}%
  \BibitemOpen
  \bibfield  {author} {\bibinfo {author} {\bibfnamefont {R.~P.}\ \bibnamefont
  {Groff}}, \bibinfo {author} {\bibfnamefont {P.}~\bibnamefont {Avakian}}, \
  and\ \bibinfo {author} {\bibfnamefont {R.~E.}\ \bibnamefont {Merrifield}},\
  }\href@noop {} {\bibfield  {journal} {\bibinfo  {journal} {Phys. Rev. B}\
  }\textbf {\bibinfo {volume} {1}},\ \bibinfo {pages} {815} (\bibinfo {year}
  {1970})}\BibitemShut {NoStop}%
\bibitem [{\citenamefont {Merrifield}(1968)}]{Merrifield68}%
  \BibitemOpen
  \bibfield  {author} {\bibinfo {author} {\bibfnamefont {R.~E.}\ \bibnamefont
  {Merrifield}},\ }\href@noop {} {\bibfield  {journal} {\bibinfo  {journal} {J.
  Chem. Phys.}\ }\textbf {\bibinfo {volume} {48}},\ \bibinfo {pages} {4318}
  (\bibinfo {year} {1968})}\BibitemShut {NoStop}%
\bibitem [{\citenamefont {Johnson}\ and\ \citenamefont
  {Merrifield}(1970)}]{Merrifield70-2}%
  \BibitemOpen
  \bibfield  {author} {\bibinfo {author} {\bibfnamefont {R.~C.}\ \bibnamefont
  {Johnson}}\ and\ \bibinfo {author} {\bibfnamefont {R.~E.}\ \bibnamefont
  {Merrifield}},\ }\href@noop {} {\bibfield  {journal} {\bibinfo  {journal}
  {Phys. Rev. B}\ }\textbf {\bibinfo {volume} {1}},\ \bibinfo {pages} {896}
  (\bibinfo {year} {1970})}\BibitemShut {NoStop}%
\bibitem [{\citenamefont {Smith}\ and\ \citenamefont {Michl}(2010)}]{Michl10}%
  \BibitemOpen
  \bibfield  {author} {\bibinfo {author} {\bibfnamefont {M.~B.}\ \bibnamefont
  {Smith}}\ and\ \bibinfo {author} {\bibfnamefont {J.}~\bibnamefont {Michl}},\
  }\href@noop {} {\bibfield  {journal} {\bibinfo  {journal} {Chem. Rev.}\
  }\textbf {\bibinfo {volume} {110}},\ \bibinfo {pages} {6891} (\bibinfo {year}
  {2010})}\BibitemShut {NoStop}%
\bibitem [{\citenamefont {Johnson}\ \emph
  {et~al.}(2013{\natexlab{a}})\citenamefont {Johnson}, \citenamefont {Nozik},\
  and\ \citenamefont {Michl}}]{Michl13-1}%
  \BibitemOpen
  \bibfield  {author} {\bibinfo {author} {\bibfnamefont {J.~C.}\ \bibnamefont
  {Johnson}}, \bibinfo {author} {\bibfnamefont {A.~J.}\ \bibnamefont {Nozik}},
  \ and\ \bibinfo {author} {\bibfnamefont {J.}~\bibnamefont {Michl}},\
  }\href@noop {} {\bibfield  {journal} {\bibinfo  {journal} {Acc. Chem. Res.}\
  }\textbf {\bibinfo {volume} {46}},\ \bibinfo {pages} {1290} (\bibinfo {year}
  {2013}{\natexlab{a}})}\BibitemShut {NoStop}%
\bibitem [{\citenamefont {Smith}\ and\ \citenamefont
  {Michl}(2013)}]{Michl13-2}%
  \BibitemOpen
  \bibfield  {author} {\bibinfo {author} {\bibfnamefont {M.~B.}\ \bibnamefont
  {Smith}}\ and\ \bibinfo {author} {\bibfnamefont {J.}~\bibnamefont {Michl}},\
  }\href@noop {} {\bibfield  {journal} {\bibinfo  {journal} {Annu. Rev. Phys.
  Chem.}\ }\textbf {\bibinfo {volume} {64}},\ \bibinfo {pages} {361} (\bibinfo
  {year} {2013})}\BibitemShut {NoStop}%
\bibitem [{\citenamefont {Chan}\ \emph {et~al.}(2011)\citenamefont {Chan},
  \citenamefont {Ligges}, \citenamefont {Jailaubekov}, \citenamefont {Kaake},
  \citenamefont {{Miaja-Avila}},\ and\ \citenamefont {Zhu}}]{Chan11}%
  \BibitemOpen
  \bibfield  {author} {\bibinfo {author} {\bibfnamefont {W.-L.}\ \bibnamefont
  {Chan}}, \bibinfo {author} {\bibfnamefont {M.}~\bibnamefont {Ligges}},
  \bibinfo {author} {\bibfnamefont {A.}~\bibnamefont {Jailaubekov}}, \bibinfo
  {author} {\bibfnamefont {L.}~\bibnamefont {Kaake}}, \bibinfo {author}
  {\bibfnamefont {L.}~\bibnamefont {{Miaja-Avila}}}, \ and\ \bibinfo {author}
  {\bibfnamefont {X.-Y.}\ \bibnamefont {Zhu}},\ }\href@noop {} {\bibfield
  {journal} {\bibinfo  {journal} {Science}\ }\textbf {\bibinfo {volume}
  {334}},\ \bibinfo {pages} {1541} (\bibinfo {year} {2011})}\BibitemShut
  {NoStop}%
\bibitem [{\citenamefont {Stern}\ \emph {et~al.}(2015)\citenamefont {Stern},
  \citenamefont {Musser}, \citenamefont {Gelinas}, \citenamefont {Parkinson},
  \citenamefont {Herz}, \citenamefont {Bruzek}, \citenamefont {Anthony},
  \citenamefont {Friend},\ and\ \citenamefont {Walker}}]{Walker15}%
  \BibitemOpen
  \bibfield  {author} {\bibinfo {author} {\bibfnamefont {H.~L.}\ \bibnamefont
  {Stern}}, \bibinfo {author} {\bibfnamefont {A.~J.}\ \bibnamefont {Musser}},
  \bibinfo {author} {\bibfnamefont {S.}~\bibnamefont {Gelinas}}, \bibinfo
  {author} {\bibfnamefont {P.}~\bibnamefont {Parkinson}}, \bibinfo {author}
  {\bibfnamefont {L.~M.}\ \bibnamefont {Herz}}, \bibinfo {author}
  {\bibfnamefont {M.~J.}\ \bibnamefont {Bruzek}}, \bibinfo {author}
  {\bibfnamefont {J.}~\bibnamefont {Anthony}}, \bibinfo {author} {\bibfnamefont
  {R.~H.}\ \bibnamefont {Friend}}, \ and\ \bibinfo {author} {\bibfnamefont
  {B.~J.}\ \bibnamefont {Walker}},\ }\href@noop {} {\bibfield  {journal}
  {\bibinfo  {journal} {Proc. Natl. Acad. Sci. U.S.A.}\ }\textbf {\bibinfo
  {volume} {112}},\ \bibinfo {pages} {7656} (\bibinfo {year}
  {2015})}\BibitemShut {NoStop}%
\bibitem [{\citenamefont {Paci}\ \emph {et~al.}(2006)\citenamefont {Paci},
  \citenamefont {Johnson}, \citenamefont {Chen}, \citenamefont {Rana},
  \citenamefont {Popovi\'{c}}, \citenamefont {David}, \citenamefont {Nozik},
  \citenamefont {Ratner},\ and\ \citenamefont {Michl}}]{Michl06}%
  \BibitemOpen
  \bibfield  {author} {\bibinfo {author} {\bibfnamefont {I.}~\bibnamefont
  {Paci}}, \bibinfo {author} {\bibfnamefont {J.~C.}\ \bibnamefont {Johnson}},
  \bibinfo {author} {\bibfnamefont {X.}~\bibnamefont {Chen}}, \bibinfo {author}
  {\bibfnamefont {G.}~\bibnamefont {Rana}}, \bibinfo {author} {\bibfnamefont
  {D.}~\bibnamefont {Popovi\'{c}}}, \bibinfo {author} {\bibfnamefont {D.~E.}\
  \bibnamefont {David}}, \bibinfo {author} {\bibfnamefont {A.~J.}\ \bibnamefont
  {Nozik}}, \bibinfo {author} {\bibfnamefont {M.~A.}\ \bibnamefont {Ratner}}, \
  and\ \bibinfo {author} {\bibfnamefont {J.}~\bibnamefont {Michl}},\
  }\href@noop {} {\bibfield  {journal} {\bibinfo  {journal} {J. Am. Chem.
  Soc.}\ }\textbf {\bibinfo {volume} {128}},\ \bibinfo {pages} {16546}
  (\bibinfo {year} {2006})}\BibitemShut {NoStop}%
\bibitem [{\citenamefont {Greyson}\ \emph
  {et~al.}(2010{\natexlab{a}})\citenamefont {Greyson}, \citenamefont {Stepp},
  \citenamefont {Chen}, \citenamefont {Schwerin}, \citenamefont {Paci},
  \citenamefont {Smith}, \citenamefont {Akdag}, \citenamefont {Johnson},
  \citenamefont {Nozik}, \citenamefont {Michl},\ and\ \citenamefont
  {Ratner}}]{Greyson10-2}%
  \BibitemOpen
  \bibfield  {author} {\bibinfo {author} {\bibfnamefont {E.~C.}\ \bibnamefont
  {Greyson}}, \bibinfo {author} {\bibfnamefont {B.~R.}\ \bibnamefont {Stepp}},
  \bibinfo {author} {\bibfnamefont {X.}~\bibnamefont {Chen}}, \bibinfo {author}
  {\bibfnamefont {A.~F.}\ \bibnamefont {Schwerin}}, \bibinfo {author}
  {\bibfnamefont {I.}~\bibnamefont {Paci}}, \bibinfo {author} {\bibfnamefont
  {M.~B.}\ \bibnamefont {Smith}}, \bibinfo {author} {\bibfnamefont
  {A.}~\bibnamefont {Akdag}}, \bibinfo {author} {\bibfnamefont {J.~C.}\
  \bibnamefont {Johnson}}, \bibinfo {author} {\bibfnamefont {A.~J.}\
  \bibnamefont {Nozik}}, \bibinfo {author} {\bibfnamefont {J.}~\bibnamefont
  {Michl}}, \ and\ \bibinfo {author} {\bibfnamefont {M.~A.}\ \bibnamefont
  {Ratner}},\ }\href@noop {} {\bibfield  {journal} {\bibinfo  {journal} {J.
  Phys. Chem. B}\ }\textbf {\bibinfo {volume} {114}},\ \bibinfo {pages} {14223}
  (\bibinfo {year} {2010}{\natexlab{a}})}\BibitemShut {NoStop}%
\bibitem [{\citenamefont {Minami}\ and\ \citenamefont
  {Nakano}(2012)}]{Nakano12-1}%
  \BibitemOpen
  \bibfield  {author} {\bibinfo {author} {\bibfnamefont {T.}~\bibnamefont
  {Minami}}\ and\ \bibinfo {author} {\bibfnamefont {M.}~\bibnamefont
  {Nakano}},\ }\href@noop {} {\bibfield  {journal} {\bibinfo  {journal} {J.
  Phys. Chem. Lett.}\ }\textbf {\bibinfo {volume} {3}},\ \bibinfo {pages} {145}
  (\bibinfo {year} {2012})}\BibitemShut {NoStop}%
\bibitem [{\citenamefont {Minami}\ \emph {et~al.}(2012)\citenamefont {Minami},
  \citenamefont {Ito},\ and\ \citenamefont {Nakano}}]{Nakano12-2}%
  \BibitemOpen
  \bibfield  {author} {\bibinfo {author} {\bibfnamefont {T.}~\bibnamefont
  {Minami}}, \bibinfo {author} {\bibfnamefont {S.}~\bibnamefont {Ito}}, \ and\
  \bibinfo {author} {\bibfnamefont {M.}~\bibnamefont {Nakano}},\ }\href@noop {}
  {\bibfield  {journal} {\bibinfo  {journal} {J. Phys. Chem. Lett.}\ }\textbf
  {\bibinfo {volume} {3}},\ \bibinfo {pages} {2719} (\bibinfo {year}
  {2012})}\BibitemShut {NoStop}%
\bibitem [{\citenamefont {Minami}\ \emph {et~al.}(2013)\citenamefont {Minami},
  \citenamefont {Ito},\ and\ \citenamefont {Nakano}}]{Nakano13}%
  \BibitemOpen
  \bibfield  {author} {\bibinfo {author} {\bibfnamefont {T.}~\bibnamefont
  {Minami}}, \bibinfo {author} {\bibfnamefont {S.}~\bibnamefont {Ito}}, \ and\
  \bibinfo {author} {\bibfnamefont {M.}~\bibnamefont {Nakano}},\ }\href@noop {}
  {\bibfield  {journal} {\bibinfo  {journal} {J. Phys. Chem. Lett.}\ }\textbf
  {\bibinfo {volume} {4}},\ \bibinfo {pages} {2133} (\bibinfo {year}
  {2013})}\BibitemShut {NoStop}%
\bibitem [{\citenamefont {Zimmerman}\ \emph {et~al.}(2010)\citenamefont
  {Zimmerman}, \citenamefont {Zhang},\ and\ \citenamefont
  {Musgrave}}]{Zimmerman10}%
  \BibitemOpen
  \bibfield  {author} {\bibinfo {author} {\bibfnamefont {P.~M.}\ \bibnamefont
  {Zimmerman}}, \bibinfo {author} {\bibfnamefont {Z.}~\bibnamefont {Zhang}}, \
  and\ \bibinfo {author} {\bibfnamefont {C.~B.}\ \bibnamefont {Musgrave}},\
  }\href@noop {} {\bibfield  {journal} {\bibinfo  {journal} {Nat. Chem.}\
  }\textbf {\bibinfo {volume} {2}},\ \bibinfo {pages} {648} (\bibinfo {year}
  {2010})}\BibitemShut {NoStop}%
\bibitem [{\citenamefont {Zimmerman}\ \emph {et~al.}(2011)\citenamefont
  {Zimmerman}, \citenamefont {Bell}, \citenamefont {Casanova},\ and\
  \citenamefont {{Head-Gordon}}}]{Zimmerman11}%
  \BibitemOpen
  \bibfield  {author} {\bibinfo {author} {\bibfnamefont {P.~M.}\ \bibnamefont
  {Zimmerman}}, \bibinfo {author} {\bibfnamefont {F.}~\bibnamefont {Bell}},
  \bibinfo {author} {\bibfnamefont {D.}~\bibnamefont {Casanova}}, \ and\
  \bibinfo {author} {\bibfnamefont {M.}~\bibnamefont {{Head-Gordon}}},\
  }\href@noop {} {\bibfield  {journal} {\bibinfo  {journal} {J. Am. Chem.
  Soc.}\ }\textbf {\bibinfo {volume} {133}},\ \bibinfo {pages} {19944}
  (\bibinfo {year} {2011})}\BibitemShut {NoStop}%
\bibitem [{\citenamefont {Zimmerman}\ \emph {et~al.}(2013)\citenamefont
  {Zimmerman}, \citenamefont {Musgrave},\ and\ \citenamefont
  {{Head-Gordon}}}]{Zimmerman12}%
  \BibitemOpen
  \bibfield  {author} {\bibinfo {author} {\bibfnamefont {P.~M.}\ \bibnamefont
  {Zimmerman}}, \bibinfo {author} {\bibfnamefont {C.~B.}\ \bibnamefont
  {Musgrave}}, \ and\ \bibinfo {author} {\bibfnamefont {M.}~\bibnamefont
  {{Head-Gordon}}},\ }\href@noop {} {\bibfield  {journal} {\bibinfo  {journal}
  {Acc. Chem. Res.}\ }\textbf {\bibinfo {volume} {46}},\ \bibinfo {pages}
  {1339} (\bibinfo {year} {2013})}\BibitemShut {NoStop}%
\bibitem [{\citenamefont {Zeng}\ \emph
  {et~al.}(2014{\natexlab{a}})\citenamefont {Zeng}, \citenamefont {Ananth},\
  and\ \citenamefont {Hoffmann}}]{Zeng14}%
  \BibitemOpen
  \bibfield  {author} {\bibinfo {author} {\bibfnamefont {T.}~\bibnamefont
  {Zeng}}, \bibinfo {author} {\bibfnamefont {N.}~\bibnamefont {Ananth}}, \ and\
  \bibinfo {author} {\bibfnamefont {R.}~\bibnamefont {Hoffmann}},\ }\href@noop
  {} {\bibfield  {journal} {\bibinfo  {journal} {J. Am. Chem. Soc.}\ }\textbf
  {\bibinfo {volume} {136}},\ \bibinfo {pages} {12638} (\bibinfo {year}
  {2014}{\natexlab{a}})}\BibitemShut {NoStop}%
\bibitem [{\citenamefont {Zeng}\ \emph
  {et~al.}(2014{\natexlab{b}})\citenamefont {Zeng}, \citenamefont {Hoffmann},\
  and\ \citenamefont {Ananth}}]{Ananth14}%
  \BibitemOpen
  \bibfield  {author} {\bibinfo {author} {\bibfnamefont {T.}~\bibnamefont
  {Zeng}}, \bibinfo {author} {\bibfnamefont {R.}~\bibnamefont {Hoffmann}}, \
  and\ \bibinfo {author} {\bibfnamefont {N.}~\bibnamefont {Ananth}},\
  }\href@noop {} {\bibfield  {journal} {\bibinfo  {journal} {J. Am. Chem.
  Soc.}\ }\textbf {\bibinfo {volume} {136}},\ \bibinfo {pages} {5755} (\bibinfo
  {year} {2014}{\natexlab{b}})}\BibitemShut {NoStop}%
\bibitem [{\citenamefont {Fuemmeler}\ \emph {et~al.}(2016)\citenamefont
  {Fuemmeler}, \citenamefont {Sanders}, \citenamefont {Pun}, \citenamefont
  {Kumarasamy}, \citenamefont {Zeng}, \citenamefont {Miyata}, \citenamefont
  {Steigerwald}, \citenamefont {Zhu}, \citenamefont {Sfeir}, \citenamefont
  {Campos},\ and\ \citenamefont {Ananth}}]{Ananth16}%
  \BibitemOpen
  \bibfield  {author} {\bibinfo {author} {\bibfnamefont {E.~G.}\ \bibnamefont
  {Fuemmeler}}, \bibinfo {author} {\bibfnamefont {S.~N.}\ \bibnamefont
  {Sanders}}, \bibinfo {author} {\bibfnamefont {A.~B.}\ \bibnamefont {Pun}},
  \bibinfo {author} {\bibfnamefont {E.}~\bibnamefont {Kumarasamy}}, \bibinfo
  {author} {\bibfnamefont {T.}~\bibnamefont {Zeng}}, \bibinfo {author}
  {\bibfnamefont {K.}~\bibnamefont {Miyata}}, \bibinfo {author} {\bibfnamefont
  {M.~L.}\ \bibnamefont {Steigerwald}}, \bibinfo {author} {\bibfnamefont
  {X.-Y.}\ \bibnamefont {Zhu}}, \bibinfo {author} {\bibfnamefont {M.~Y.}\
  \bibnamefont {Sfeir}}, \bibinfo {author} {\bibfnamefont {L.~M.}\ \bibnamefont
  {Campos}}, \ and\ \bibinfo {author} {\bibfnamefont {N.}~\bibnamefont
  {Ananth}},\ }\href@noop {} {\bibfield  {journal} {\bibinfo  {journal} {ACS
  Cent. Sci.}\ }\textbf {\bibinfo {volume} {2}},\ \bibinfo {pages} {316}
  (\bibinfo {year} {2016})}\BibitemShut {NoStop}%
\bibitem [{\citenamefont {Havenith}\ \emph {et~al.}(2012)\citenamefont
  {Havenith}, \citenamefont {{de Gier}},\ and\ \citenamefont
  {Broer}}]{deGier12}%
  \BibitemOpen
  \bibfield  {author} {\bibinfo {author} {\bibfnamefont {R.~W.~A.}\
  \bibnamefont {Havenith}}, \bibinfo {author} {\bibfnamefont {H.~D.}\
  \bibnamefont {{de Gier}}}, \ and\ \bibinfo {author} {\bibfnamefont
  {R.}~\bibnamefont {Broer}},\ }\href@noop {} {\bibfield  {journal} {\bibinfo
  {journal} {Mol. Phys.}\ }\textbf {\bibinfo {volume} {110}},\ \bibinfo {pages}
  {2445} (\bibinfo {year} {2012})}\BibitemShut {NoStop}%
\bibitem [{\citenamefont {Parker}\ \emph {et~al.}(2014)\citenamefont {Parker},
  \citenamefont {Seideman}, \citenamefont {Ratner},\ and\ \citenamefont
  {Shiozaki}}]{Parker14}%
  \BibitemOpen
  \bibfield  {author} {\bibinfo {author} {\bibfnamefont {S.~M.}\ \bibnamefont
  {Parker}}, \bibinfo {author} {\bibfnamefont {T.}~\bibnamefont {Seideman}},
  \bibinfo {author} {\bibfnamefont {M.~A.}\ \bibnamefont {Ratner}}, \ and\
  \bibinfo {author} {\bibfnamefont {T.}~\bibnamefont {Shiozaki}},\ }\href@noop
  {} {\bibfield  {journal} {\bibinfo  {journal} {J. Phys. Chem. C}\ }\textbf
  {\bibinfo {volume} {118}},\ \bibinfo {pages} {12700} (\bibinfo {year}
  {2014})}\BibitemShut {NoStop}%
\bibitem [{\citenamefont {Aryanpour}\ \emph
  {et~al.}(2015{\natexlab{a}})\citenamefont {Aryanpour}, \citenamefont
  {Shukla},\ and\ \citenamefont {Mazumdar}}]{Aryanpour15-1}%
  \BibitemOpen
  \bibfield  {author} {\bibinfo {author} {\bibfnamefont {K.}~\bibnamefont
  {Aryanpour}}, \bibinfo {author} {\bibfnamefont {A.}~\bibnamefont {Shukla}}, \
  and\ \bibinfo {author} {\bibfnamefont {S.}~\bibnamefont {Mazumdar}},\
  }\href@noop {} {\bibfield  {journal} {\bibinfo  {journal} {J. Phys. Chem. C}\
  }\textbf {\bibinfo {volume} {119}},\ \bibinfo {pages} {6966} (\bibinfo {year}
  {2015}{\natexlab{a}})}\BibitemShut {NoStop}%
\bibitem [{\citenamefont {Gradinaru}\ \emph {et~al.}(2001)\citenamefont
  {Gradinaru}, \citenamefont {Kennis}, \citenamefont {Papagiannakis},
  \citenamefont {{van Stokkum}}, \citenamefont {Cogdell}, \citenamefont
  {Fleming}, \citenamefont {Niederman},\ and\ \citenamefont {{van
  Grondelle}}}]{Gradinaru01}%
  \BibitemOpen
  \bibfield  {author} {\bibinfo {author} {\bibfnamefont {C.~C.}\ \bibnamefont
  {Gradinaru}}, \bibinfo {author} {\bibfnamefont {J.~T.~M.}\ \bibnamefont
  {Kennis}}, \bibinfo {author} {\bibfnamefont {E.}~\bibnamefont
  {Papagiannakis}}, \bibinfo {author} {\bibfnamefont {I.~H.~M.}\ \bibnamefont
  {{van Stokkum}}}, \bibinfo {author} {\bibfnamefont {R.~J.}\ \bibnamefont
  {Cogdell}}, \bibinfo {author} {\bibfnamefont {G.~R.}\ \bibnamefont
  {Fleming}}, \bibinfo {author} {\bibfnamefont {R.~A.}\ \bibnamefont
  {Niederman}}, \ and\ \bibinfo {author} {\bibfnamefont {R.}~\bibnamefont {{van
  Grondelle}}},\ }\href@noop {} {\bibfield  {journal} {\bibinfo  {journal}
  {Proc. Natl. Acad. Sci. U.S.A.}\ }\textbf {\bibinfo {volume} {98}},\ \bibinfo
  {pages} {2364} (\bibinfo {year} {2001})}\BibitemShut {NoStop}%
\bibitem [{\citenamefont {Papagiannakis}\ \emph {et~al.}(2002)\citenamefont
  {Papagiannakis}, \citenamefont {Kennis}, \citenamefont {{van Stokkum}},
  \citenamefont {Cogdell},\ and\ \citenamefont {{van Grondelle}}}]{Kennis02}%
  \BibitemOpen
  \bibfield  {author} {\bibinfo {author} {\bibfnamefont {E.}~\bibnamefont
  {Papagiannakis}}, \bibinfo {author} {\bibfnamefont {J.~T.~M.}\ \bibnamefont
  {Kennis}}, \bibinfo {author} {\bibfnamefont {I.~H.~M.}\ \bibnamefont {{van
  Stokkum}}}, \bibinfo {author} {\bibfnamefont {R.~J.}\ \bibnamefont
  {Cogdell}}, \ and\ \bibinfo {author} {\bibfnamefont {R.}~\bibnamefont {{van
  Grondelle}}},\ }\href@noop {} {\bibfield  {journal} {\bibinfo  {journal}
  {Proc. Natl. Acad. Sci. U.S.A.}\ }\textbf {\bibinfo {volume} {99}},\ \bibinfo
  {pages} {6017} (\bibinfo {year} {2002})}\BibitemShut {NoStop}%
\bibitem [{\citenamefont {Wang}\ and\ \citenamefont {Tauber}(2010)}]{Tauber10}%
  \BibitemOpen
  \bibfield  {author} {\bibinfo {author} {\bibfnamefont {C.}~\bibnamefont
  {Wang}}\ and\ \bibinfo {author} {\bibfnamefont {M.~J.}\ \bibnamefont
  {Tauber}},\ }\href@noop {} {\bibfield  {journal} {\bibinfo  {journal} {J. Am.
  Chem. Soc.}\ }\textbf {\bibinfo {volume} {132}},\ \bibinfo {pages} {13988}
  (\bibinfo {year} {2010})}\BibitemShut {NoStop}%
\bibitem [{\citenamefont {Wang}\ \emph {et~al.}(2012)\citenamefont {Wang},
  \citenamefont {Angelella}, \citenamefont {Kuo},\ and\ \citenamefont
  {Tauber}}]{Tauber12}%
  \BibitemOpen
  \bibfield  {author} {\bibinfo {author} {\bibfnamefont {C.}~\bibnamefont
  {Wang}}, \bibinfo {author} {\bibfnamefont {M.}~\bibnamefont {Angelella}},
  \bibinfo {author} {\bibfnamefont {C.-H.}\ \bibnamefont {Kuo}}, \ and\
  \bibinfo {author} {\bibfnamefont {M.~J.}\ \bibnamefont {Tauber}},\
  }\href@noop {} {\bibfield  {journal} {\bibinfo  {journal} {Proc. SPIE}\
  }\textbf {\bibinfo {volume} {8459}},\ \bibinfo {pages} {845905} (\bibinfo
  {year} {2012})}\BibitemShut {NoStop}%
\bibitem [{\citenamefont {Dillon}\ \emph {et~al.}(2013)\citenamefont {Dillon},
  \citenamefont {Piland},\ and\ \citenamefont {Bardeen}}]{Bardeen13-2}%
  \BibitemOpen
  \bibfield  {author} {\bibinfo {author} {\bibfnamefont {R.~J.}\ \bibnamefont
  {Dillon}}, \bibinfo {author} {\bibfnamefont {G.~B.}\ \bibnamefont {Piland}},
  \ and\ \bibinfo {author} {\bibfnamefont {C.~J.}\ \bibnamefont {Bardeen}},\
  }\href@noop {} {\bibfield  {journal} {\bibinfo  {journal} {J. Am. Chem.
  Soc.}\ }\textbf {\bibinfo {volume} {135}},\ \bibinfo {pages} {17278}
  (\bibinfo {year} {2013})}\BibitemShut {NoStop}%
\bibitem [{\citenamefont {Musser}\ \emph {et~al.}(2013)\citenamefont {Musser},
  \citenamefont {{Al-Hashimi}}, \citenamefont {Maiuri}, \citenamefont {Brida},
  \citenamefont {Heeney}, \citenamefont {Cerullo}, \citenamefont {Friend},\
  and\ \citenamefont {Clark}}]{Musser13}%
  \BibitemOpen
  \bibfield  {author} {\bibinfo {author} {\bibfnamefont {A.~J.}\ \bibnamefont
  {Musser}}, \bibinfo {author} {\bibfnamefont {M.}~\bibnamefont
  {{Al-Hashimi}}}, \bibinfo {author} {\bibfnamefont {M.}~\bibnamefont
  {Maiuri}}, \bibinfo {author} {\bibfnamefont {D.}~\bibnamefont {Brida}},
  \bibinfo {author} {\bibfnamefont {M.}~\bibnamefont {Heeney}}, \bibinfo
  {author} {\bibfnamefont {G.}~\bibnamefont {Cerullo}}, \bibinfo {author}
  {\bibfnamefont {R.~H.}\ \bibnamefont {Friend}}, \ and\ \bibinfo {author}
  {\bibfnamefont {J.}~\bibnamefont {Clark}},\ }\href@noop {} {\bibfield
  {journal} {\bibinfo  {journal} {J. Am. Chem. Soc.}\ }\textbf {\bibinfo
  {volume} {135}},\ \bibinfo {pages} {12747} (\bibinfo {year}
  {2013})}\BibitemShut {NoStop}%
\bibitem [{\citenamefont {Musser}\ \emph {et~al.}(2015)\citenamefont {Musser},
  \citenamefont {Maiuri}, \citenamefont {Brida}, \citenamefont {Cerullo},
  \citenamefont {Friend},\ and\ \citenamefont {Clark}}]{Musser15}%
  \BibitemOpen
  \bibfield  {author} {\bibinfo {author} {\bibfnamefont {A.~J.}\ \bibnamefont
  {Musser}}, \bibinfo {author} {\bibfnamefont {M.}~\bibnamefont {Maiuri}},
  \bibinfo {author} {\bibfnamefont {D.}~\bibnamefont {Brida}}, \bibinfo
  {author} {\bibfnamefont {G.}~\bibnamefont {Cerullo}}, \bibinfo {author}
  {\bibfnamefont {R.~H.}\ \bibnamefont {Friend}}, \ and\ \bibinfo {author}
  {\bibfnamefont {J.}~\bibnamefont {Clark}},\ }\href@noop {} {\bibfield
  {journal} {\bibinfo  {journal} {J. Am. Chem. Soc.}\ }\textbf {\bibinfo
  {volume} {137}},\ \bibinfo {pages} {5130} (\bibinfo {year}
  {2015})}\BibitemShut {NoStop}%
\bibitem [{\citenamefont {Feng}\ \emph {et~al.}(2013)\citenamefont {Feng},
  \citenamefont {Luzanov},\ and\ \citenamefont {Krylov}}]{Krylov13}%
  \BibitemOpen
  \bibfield  {author} {\bibinfo {author} {\bibfnamefont {X.}~\bibnamefont
  {Feng}}, \bibinfo {author} {\bibfnamefont {A.~V.}\ \bibnamefont {Luzanov}}, \
  and\ \bibinfo {author} {\bibfnamefont {A.~I.}\ \bibnamefont {Krylov}},\
  }\href@noop {} {\bibfield  {journal} {\bibinfo  {journal} {J. Phys. Chem.
  Lett.}\ }\textbf {\bibinfo {volume} {4}},\ \bibinfo {pages} {3845} (\bibinfo
  {year} {2013})}\BibitemShut {NoStop}%
\bibitem [{\citenamefont {Renaud}\ \emph {et~al.}(2013)\citenamefont {Renaud},
  \citenamefont {Sherratt},\ and\ \citenamefont {Ratner}}]{Ratner13}%
  \BibitemOpen
  \bibfield  {author} {\bibinfo {author} {\bibfnamefont {N.}~\bibnamefont
  {Renaud}}, \bibinfo {author} {\bibfnamefont {P.~A.}\ \bibnamefont
  {Sherratt}}, \ and\ \bibinfo {author} {\bibfnamefont {M.~A.}\ \bibnamefont
  {Ratner}},\ }\href@noop {} {\bibfield  {journal} {\bibinfo  {journal} {J.
  Phys. Chem. Lett.}\ }\textbf {\bibinfo {volume} {4}},\ \bibinfo {pages}
  {1065} (\bibinfo {year} {2013})}\BibitemShut {NoStop}%
\bibitem [{\citenamefont {Eaton}\ \emph {et~al.}(2013)\citenamefont {Eaton},
  \citenamefont {Shoer}, \citenamefont {Karlen}, \citenamefont {Dyar},
  \citenamefont {Margulies}, \citenamefont {Veldkamp}, \citenamefont {Ramanan},
  \citenamefont {Hartzler}, \citenamefont {Savikhin}, \citenamefont {Marks},\
  and\ \citenamefont {Wasielewski}}]{Wasielewski13}%
  \BibitemOpen
  \bibfield  {author} {\bibinfo {author} {\bibfnamefont {S.~W.}\ \bibnamefont
  {Eaton}}, \bibinfo {author} {\bibfnamefont {L.~E.}\ \bibnamefont {Shoer}},
  \bibinfo {author} {\bibfnamefont {S.~D.}\ \bibnamefont {Karlen}}, \bibinfo
  {author} {\bibfnamefont {S.~M.}\ \bibnamefont {Dyar}}, \bibinfo {author}
  {\bibfnamefont {E.~A.}\ \bibnamefont {Margulies}}, \bibinfo {author}
  {\bibfnamefont {B.~S.}\ \bibnamefont {Veldkamp}}, \bibinfo {author}
  {\bibfnamefont {C.}~\bibnamefont {Ramanan}}, \bibinfo {author} {\bibfnamefont
  {D.~A.}\ \bibnamefont {Hartzler}}, \bibinfo {author} {\bibfnamefont
  {S.}~\bibnamefont {Savikhin}}, \bibinfo {author} {\bibfnamefont {T.~J.}\
  \bibnamefont {Marks}}, \ and\ \bibinfo {author} {\bibfnamefont {M.~R.}\
  \bibnamefont {Wasielewski}},\ }\href@noop {} {\bibfield  {journal} {\bibinfo
  {journal} {J. Am. Chem. Soc.}\ }\textbf {\bibinfo {volume} {135}},\ \bibinfo
  {pages} {14701−14712} (\bibinfo {year} {2013})}\BibitemShut {NoStop}%
\bibitem [{\citenamefont {Wang}\ \emph {et~al.}(2014)\citenamefont {Wang},
  \citenamefont {Olivier}, \citenamefont {Prezhdo},\ and\ \citenamefont
  {Beljonne}}]{Beljonne14}%
  \BibitemOpen
  \bibfield  {author} {\bibinfo {author} {\bibfnamefont {L.}~\bibnamefont
  {Wang}}, \bibinfo {author} {\bibfnamefont {Y.}~\bibnamefont {Olivier}},
  \bibinfo {author} {\bibfnamefont {O.~V.}\ \bibnamefont {Prezhdo}}, \ and\
  \bibinfo {author} {\bibfnamefont {D.}~\bibnamefont {Beljonne}},\ }\href@noop
  {} {\bibfield  {journal} {\bibinfo  {journal} {J. Phys. Chem. Lett.}\
  }\textbf {\bibinfo {volume} {5}},\ \bibinfo {pages} {3345} (\bibinfo {year}
  {2014})}\BibitemShut {NoStop}%
\bibitem [{\citenamefont {Renaud}\ and\ \citenamefont
  {Grozema}(2015)}]{Grozema15}%
  \BibitemOpen
  \bibfield  {author} {\bibinfo {author} {\bibfnamefont {N.}~\bibnamefont
  {Renaud}}\ and\ \bibinfo {author} {\bibfnamefont {F.~C.}\ \bibnamefont
  {Grozema}},\ }\href@noop {} {\bibfield  {journal} {\bibinfo  {journal} {J.
  Phys. Chem. Lett.}\ }\textbf {\bibinfo {volume} {6}},\ \bibinfo {pages} {360}
  (\bibinfo {year} {2015})}\BibitemShut {NoStop}%
\bibitem [{\citenamefont {Walker}\ \emph {et~al.}(2013)\citenamefont {Walker},
  \citenamefont {Musser}, \citenamefont {Beljonne},\ and\ \citenamefont
  {Friend}}]{Friend13}%
  \BibitemOpen
  \bibfield  {author} {\bibinfo {author} {\bibfnamefont {B.~J.}\ \bibnamefont
  {Walker}}, \bibinfo {author} {\bibfnamefont {A.~J.}\ \bibnamefont {Musser}},
  \bibinfo {author} {\bibfnamefont {D.}~\bibnamefont {Beljonne}}, \ and\
  \bibinfo {author} {\bibfnamefont {R.~H.}\ \bibnamefont {Friend}},\
  }\href@noop {} {\bibfield  {journal} {\bibinfo  {journal} {Nat. Chem.}\
  }\textbf {\bibinfo {volume} {5}},\ \bibinfo {pages} {1019} (\bibinfo {year}
  {2013})}\BibitemShut {NoStop}%
\bibitem [{\citenamefont {Zirzlmeier}\ \emph {et~al.}(2013)\citenamefont
  {Zirzlmeier}, \citenamefont {Lehnherr}, \citenamefont {Coto}, \citenamefont
  {Chernick}, \citenamefont {Casillas}, \citenamefont {Basel}, \citenamefont
  {Thoss}, \citenamefont {Tykwinskid},\ and\ \citenamefont {Guldi}}]{Guldi15}%
  \BibitemOpen
  \bibfield  {author} {\bibinfo {author} {\bibfnamefont {J.}~\bibnamefont
  {Zirzlmeier}}, \bibinfo {author} {\bibfnamefont {D.}~\bibnamefont
  {Lehnherr}}, \bibinfo {author} {\bibfnamefont {P.~B.}\ \bibnamefont {Coto}},
  \bibinfo {author} {\bibfnamefont {E.~T.}\ \bibnamefont {Chernick}}, \bibinfo
  {author} {\bibfnamefont {R.}~\bibnamefont {Casillas}}, \bibinfo {author}
  {\bibfnamefont {B.~S.}\ \bibnamefont {Basel}}, \bibinfo {author}
  {\bibfnamefont {M.}~\bibnamefont {Thoss}}, \bibinfo {author} {\bibfnamefont
  {R.~R.}\ \bibnamefont {Tykwinskid}}, \ and\ \bibinfo {author} {\bibfnamefont
  {D.~M.}\ \bibnamefont {Guldi}},\ }\href@noop {} {\bibfield  {journal}
  {\bibinfo  {journal} {Proc. Natl. Acad. Sci. U.S.A.}\ }\textbf {\bibinfo
  {volume} {112}},\ \bibinfo {pages} {5325} (\bibinfo {year}
  {2013})}\BibitemShut {NoStop}%
\bibitem [{\citenamefont {Sanders}\ \emph {et~al.}(2015)\citenamefont
  {Sanders}, \citenamefont {Kumarasamy}, \citenamefont {Pun}, \citenamefont
  {Trinh}, \citenamefont {Choi}, \citenamefont {Xia}, \citenamefont {Taffet},
  \citenamefont {Low}, \citenamefont {Miller}, \citenamefont {Roy},
  \citenamefont {Zhu}, \citenamefont {Steigerwald}, \citenamefont {Sfeir},\
  and\ \citenamefont {Campos}}]{Steigerwald15}%
  \BibitemOpen
  \bibfield  {author} {\bibinfo {author} {\bibfnamefont {S.~N.}\ \bibnamefont
  {Sanders}}, \bibinfo {author} {\bibfnamefont {E.}~\bibnamefont {Kumarasamy}},
  \bibinfo {author} {\bibfnamefont {A.~B.}\ \bibnamefont {Pun}}, \bibinfo
  {author} {\bibfnamefont {M.~T.}\ \bibnamefont {Trinh}}, \bibinfo {author}
  {\bibfnamefont {B.}~\bibnamefont {Choi}}, \bibinfo {author} {\bibfnamefont
  {J.}~\bibnamefont {Xia}}, \bibinfo {author} {\bibfnamefont {E.~J.}\
  \bibnamefont {Taffet}}, \bibinfo {author} {\bibfnamefont {J.~Z.}\
  \bibnamefont {Low}}, \bibinfo {author} {\bibfnamefont {J.~R.}\ \bibnamefont
  {Miller}}, \bibinfo {author} {\bibfnamefont {X.}~\bibnamefont {Roy}},
  \bibinfo {author} {\bibfnamefont {X.-Y.}\ \bibnamefont {Zhu}}, \bibinfo
  {author} {\bibfnamefont {M.~L.}\ \bibnamefont {Steigerwald}}, \bibinfo
  {author} {\bibfnamefont {M.~Y.}\ \bibnamefont {Sfeir}}, \ and\ \bibinfo
  {author} {\bibfnamefont {L.~M.}\ \bibnamefont {Campos}},\ }\href@noop {}
  {\bibfield  {journal} {\bibinfo  {journal} {J. Am. Chem. Soc.}\ }\textbf
  {\bibinfo {volume} {137}},\ \bibinfo {pages} {8965} (\bibinfo {year}
  {2015})}\BibitemShut {NoStop}%
\bibitem [{\citenamefont {Hudson}\ and\ \citenamefont
  {Kohler}(1972)}]{Hudson72}%
  \BibitemOpen
  \bibfield  {author} {\bibinfo {author} {\bibfnamefont {B.~S.}\ \bibnamefont
  {Hudson}}\ and\ \bibinfo {author} {\bibfnamefont {B.~E.}\ \bibnamefont
  {Kohler}},\ }\href@noop {} {\bibfield  {journal} {\bibinfo  {journal} {Chem.
  Phys. Lett.}\ }\textbf {\bibinfo {volume} {14}},\ \bibinfo {pages} {299}
  (\bibinfo {year} {1972})}\BibitemShut {NoStop}%
\bibitem [{\citenamefont {Schulten}\ and\ \citenamefont
  {Karplus}(1972)}]{Schulten72}%
  \BibitemOpen
  \bibfield  {author} {\bibinfo {author} {\bibfnamefont {K.}~\bibnamefont
  {Schulten}}\ and\ \bibinfo {author} {\bibfnamefont {M.}~\bibnamefont
  {Karplus}},\ }\href@noop {} {\bibfield  {journal} {\bibinfo  {journal} {Chem.
  Phys. Lett.}\ }\textbf {\bibinfo {volume} {14}},\ \bibinfo {pages} {305}
  (\bibinfo {year} {1972})}\BibitemShut {NoStop}%
\bibitem [{\citenamefont {Tavan}\ and\ \citenamefont
  {Schulten}(1987)}]{Schulten87}%
  \BibitemOpen
  \bibfield  {author} {\bibinfo {author} {\bibfnamefont {P.}~\bibnamefont
  {Tavan}}\ and\ \bibinfo {author} {\bibfnamefont {K.}~\bibnamefont
  {Schulten}},\ }\href@noop {} {\bibfield  {journal} {\bibinfo  {journal}
  {Phys. Rev. B}\ }\textbf {\bibinfo {volume} {36}},\ \bibinfo {pages} {4337}
  (\bibinfo {year} {1987})}\BibitemShut {NoStop}%
\bibitem [{\citenamefont {Greyson}\ \emph
  {et~al.}(2010{\natexlab{b}})\citenamefont {Greyson}, \citenamefont
  {{Vura-Weis}}, \citenamefont {Michl},\ and\ \citenamefont
  {Ratner}}]{Greyson10-1}%
  \BibitemOpen
  \bibfield  {author} {\bibinfo {author} {\bibfnamefont {E.~C.}\ \bibnamefont
  {Greyson}}, \bibinfo {author} {\bibfnamefont {J.}~\bibnamefont
  {{Vura-Weis}}}, \bibinfo {author} {\bibfnamefont {J.}~\bibnamefont {Michl}},
  \ and\ \bibinfo {author} {\bibfnamefont {M.~A.}\ \bibnamefont {Ratner}},\
  }\href@noop {} {\bibfield  {journal} {\bibinfo  {journal} {J. Phys. Chem. B}\
  }\textbf {\bibinfo {volume} {114}},\ \bibinfo {pages} {14168} (\bibinfo
  {year} {2010}{\natexlab{b}})}\BibitemShut {NoStop}%
\bibitem [{\citenamefont {Chen}\ \emph {et~al.}(2014)\citenamefont {Chen},
  \citenamefont {Shen},\ and\ \citenamefont {Li}}]{Chen14}%
  \BibitemOpen
  \bibfield  {author} {\bibinfo {author} {\bibfnamefont {Y.}~\bibnamefont
  {Chen}}, \bibinfo {author} {\bibfnamefont {L.}~\bibnamefont {Shen}}, \ and\
  \bibinfo {author} {\bibfnamefont {X.}~\bibnamefont {Li}},\ }\href@noop {}
  {\bibfield  {journal} {\bibinfo  {journal} {J. Phys. Chem. A}\ }\textbf
  {\bibinfo {volume} {118}},\ \bibinfo {pages} {5700} (\bibinfo {year}
  {2014})}\BibitemShut {NoStop}%
\bibitem [{\citenamefont {Busby}\ \emph {et~al.}(2015)\citenamefont {Busby},
  \citenamefont {Xia}, \citenamefont {Wu}, \citenamefont {Low}, \citenamefont
  {Song}, \citenamefont {Miller}, \citenamefont {Zhu}, \citenamefont {Campos},\
  and\ \citenamefont {Sfeir}}]{Sfeir15}%
  \BibitemOpen
  \bibfield  {author} {\bibinfo {author} {\bibfnamefont {E.}~\bibnamefont
  {Busby}}, \bibinfo {author} {\bibfnamefont {J.}~\bibnamefont {Xia}}, \bibinfo
  {author} {\bibfnamefont {Q.}~\bibnamefont {Wu}}, \bibinfo {author}
  {\bibfnamefont {J.~Z.}\ \bibnamefont {Low}}, \bibinfo {author} {\bibfnamefont
  {R.}~\bibnamefont {Song}}, \bibinfo {author} {\bibfnamefont {J.~R.}\
  \bibnamefont {Miller}}, \bibinfo {author} {\bibfnamefont {X.-Y.}\
  \bibnamefont {Zhu}}, \bibinfo {author} {\bibfnamefont {L.~M.}\ \bibnamefont
  {Campos}}, \ and\ \bibinfo {author} {\bibfnamefont {M.~Y.}\ \bibnamefont
  {Sfeir}},\ }\href@noop {} {\bibfield  {journal} {\bibinfo  {journal} {Nat.
  Mater.}\ }\textbf {\bibinfo {volume} {14}},\ \bibinfo {pages} {426} (\bibinfo
  {year} {2015})}\BibitemShut {NoStop}%
\bibitem [{\citenamefont {Aryanpour}\ \emph
  {et~al.}(2015{\natexlab{b}})\citenamefont {Aryanpour}, \citenamefont {Dutta},
  \citenamefont {Huynh}, \citenamefont {Vardeny},\ and\ \citenamefont
  {Mazumdar}}]{Aryanpour15-2}%
  \BibitemOpen
  \bibfield  {author} {\bibinfo {author} {\bibfnamefont {K.}~\bibnamefont
  {Aryanpour}}, \bibinfo {author} {\bibfnamefont {T.}~\bibnamefont {Dutta}},
  \bibinfo {author} {\bibfnamefont {U.~N.~V.}\ \bibnamefont {Huynh}}, \bibinfo
  {author} {\bibfnamefont {Z.~V.}\ \bibnamefont {Vardeny}}, \ and\ \bibinfo
  {author} {\bibfnamefont {S.}~\bibnamefont {Mazumdar}},\ }\href@noop {}
  {\bibfield  {journal} {\bibinfo  {journal} {Phys. Rev. Lett.}\ }\textbf
  {\bibinfo {volume} {115}},\ \bibinfo {pages} {267401} (\bibinfo {year}
  {2015}{\natexlab{b}})}\BibitemShut {NoStop}%
\bibitem [{\citenamefont {Kasai}\ \emph {et~al.}(2015)\citenamefont {Kasai},
  \citenamefont {Tamai}, \citenamefont {Ohkita}, \citenamefont {Benten},\ and\
  \citenamefont {Ito}}]{Kasai15}%
  \BibitemOpen
  \bibfield  {author} {\bibinfo {author} {\bibfnamefont {Y.}~\bibnamefont
  {Kasai}}, \bibinfo {author} {\bibfnamefont {Y.}~\bibnamefont {Tamai}},
  \bibinfo {author} {\bibfnamefont {H.}~\bibnamefont {Ohkita}}, \bibinfo
  {author} {\bibfnamefont {H.}~\bibnamefont {Benten}}, \ and\ \bibinfo {author}
  {\bibfnamefont {S.}~\bibnamefont {Ito}},\ }\href@noop {} {\bibfield
  {journal} {\bibinfo  {journal} {J. Am. Chem. Soc.}\ }\textbf {\bibinfo
  {volume} {137}},\ \bibinfo {pages} {15980} (\bibinfo {year}
  {2015})}\BibitemShut {NoStop}%
\bibitem [{\citenamefont {{M\"{u}ller}}\ \emph {et~al.}(2006)\citenamefont
  {{M\"{u}ller}}, \citenamefont {Avlasevich}, \citenamefont {{M\"{u}llen}},\
  and\ \citenamefont {Bardeen}}]{Bardeen06}%
  \BibitemOpen
  \bibfield  {author} {\bibinfo {author} {\bibfnamefont {A.~M.}\ \bibnamefont
  {{M\"{u}ller}}}, \bibinfo {author} {\bibfnamefont {Y.~S.}\ \bibnamefont
  {Avlasevich}}, \bibinfo {author} {\bibfnamefont {K.}~\bibnamefont
  {{M\"{u}llen}}}, \ and\ \bibinfo {author} {\bibfnamefont {C.~J.}\
  \bibnamefont {Bardeen}},\ }\href@noop {} {\bibfield  {journal} {\bibinfo
  {journal} {Chem. Phys. Lett.}\ }\textbf {\bibinfo {volume} {421}},\ \bibinfo
  {pages} {518} (\bibinfo {year} {2006})}\BibitemShut {NoStop}%
\bibitem [{\citenamefont {{M\"{u}ller}}\ \emph {et~al.}(2007)\citenamefont
  {{M\"{u}ller}}, \citenamefont {Avlasevich}, \citenamefont {Schoeller},
  \citenamefont {{M\"{u}llen}},\ and\ \citenamefont {Bardeen}}]{Bardeen07}%
  \BibitemOpen
  \bibfield  {author} {\bibinfo {author} {\bibfnamefont {A.~M.}\ \bibnamefont
  {{M\"{u}ller}}}, \bibinfo {author} {\bibfnamefont {Y.~S.}\ \bibnamefont
  {Avlasevich}}, \bibinfo {author} {\bibfnamefont {W.~W.}\ \bibnamefont
  {Schoeller}}, \bibinfo {author} {\bibfnamefont {K.}~\bibnamefont
  {{M\"{u}llen}}}, \ and\ \bibinfo {author} {\bibfnamefont {C.~J.}\
  \bibnamefont {Bardeen}},\ }\href@noop {} {\bibfield  {journal} {\bibinfo
  {journal} {J. Am. Chem. Soc.}\ }\textbf {\bibinfo {volume} {129}},\ \bibinfo
  {pages} {14240} (\bibinfo {year} {2007})}\BibitemShut {NoStop}%
\bibitem [{\citenamefont {Vallett}\ \emph {et~al.}(2013)\citenamefont
  {Vallett}, \citenamefont {Snyder},\ and\ \citenamefont
  {Damrauer}}]{Damrauer13}%
  \BibitemOpen
  \bibfield  {author} {\bibinfo {author} {\bibfnamefont {P.~J.}\ \bibnamefont
  {Vallett}}, \bibinfo {author} {\bibfnamefont {J.~L.}\ \bibnamefont {Snyder}},
  \ and\ \bibinfo {author} {\bibfnamefont {N.~H.}\ \bibnamefont {Damrauer}},\
  }\href@noop {} {\bibfield  {journal} {\bibinfo  {journal} {J. Phys. Chem. A}\
  }\textbf {\bibinfo {volume} {117}},\ \bibinfo {pages} {10824} (\bibinfo
  {year} {2013})}\BibitemShut {NoStop}%
\bibitem [{\citenamefont {Korovina}\ \emph {et~al.}(2016)\citenamefont
  {Korovina}, \citenamefont {Das}, \citenamefont {Nett}, \citenamefont {Feng},
  \citenamefont {Joy}, \citenamefont {Haiges}, \citenamefont {Krylov},
  \citenamefont {Bradforth},\ and\ \citenamefont {Thompson}}]{Thompson16}%
  \BibitemOpen
  \bibfield  {author} {\bibinfo {author} {\bibfnamefont {N.~V.}\ \bibnamefont
  {Korovina}}, \bibinfo {author} {\bibfnamefont {S.}~\bibnamefont {Das}},
  \bibinfo {author} {\bibfnamefont {Z.}~\bibnamefont {Nett}}, \bibinfo {author}
  {\bibfnamefont {X.}~\bibnamefont {Feng}}, \bibinfo {author} {\bibfnamefont
  {J.}~\bibnamefont {Joy}}, \bibinfo {author} {\bibfnamefont {R.}~\bibnamefont
  {Haiges}}, \bibinfo {author} {\bibfnamefont {A.~I.}\ \bibnamefont {Krylov}},
  \bibinfo {author} {\bibfnamefont {S.~E.}\ \bibnamefont {Bradforth}}, \ and\
  \bibinfo {author} {\bibfnamefont {M.~E.}\ \bibnamefont {Thompson}},\
  }\href@noop {} {\bibfield  {journal} {\bibinfo  {journal} {J. Am. Chem.
  Soc.}\ }\textbf {\bibinfo {volume} {138}},\ \bibinfo {pages} {617} (\bibinfo
  {year} {2016})}\BibitemShut {NoStop}%
\bibitem [{\citenamefont {Feng}\ \emph {et~al.}(2016)\citenamefont {Feng},
  \citenamefont {Casanova},\ and\ \citenamefont {Krylov}}]{Krylov16}%
  \BibitemOpen
  \bibfield  {author} {\bibinfo {author} {\bibfnamefont {X.}~\bibnamefont
  {Feng}}, \bibinfo {author} {\bibfnamefont {D.}~\bibnamefont {Casanova}}, \
  and\ \bibinfo {author} {\bibfnamefont {A.~I.}\ \bibnamefont {Krylov}},\
  }\href@noop {} {\bibfield  {journal} {\bibinfo  {journal} {J. Phys. Chem. C}\
  }\textbf {\bibinfo {volume} {120}},\ \bibinfo {pages} {19070} (\bibinfo
  {year} {2016})}\BibitemShut {NoStop}%
\bibitem [{\citenamefont {Cook}\ \emph {et~al.}(2016)\citenamefont {Cook},
  \citenamefont {Carey},\ and\ \citenamefont {Damrauer}}]{Damrauer16}%
  \BibitemOpen
  \bibfield  {author} {\bibinfo {author} {\bibfnamefont {J.~D.}\ \bibnamefont
  {Cook}}, \bibinfo {author} {\bibfnamefont {T.~J.}\ \bibnamefont {Carey}}, \
  and\ \bibinfo {author} {\bibfnamefont {N.~H.}\ \bibnamefont {Damrauer}},\
  }\href@noop {} {\bibfield  {journal} {\bibinfo  {journal} {J. Phys. Chem. A}\
  }\textbf {\bibinfo {volume} {120}},\ \bibinfo {pages} {4473} (\bibinfo {year}
  {2016})}\BibitemShut {NoStop}%
\bibitem [{\citenamefont {Lukman}\ \emph {et~al.}(2015)\citenamefont {Lukman},
  \citenamefont {Musser}, \citenamefont {Chen}, \citenamefont {Athanasopoulos},
  \citenamefont {Yong}, \citenamefont {Zeng}, \citenamefont {Ye}, \citenamefont
  {Chi}, \citenamefont {Hodgkiss}, \citenamefont {Wu}, \citenamefont {Friend},
  ,\ and\ \citenamefont {Greenham}}]{Greenham15}%
  \BibitemOpen
  \bibfield  {author} {\bibinfo {author} {\bibfnamefont {S.}~\bibnamefont
  {Lukman}}, \bibinfo {author} {\bibfnamefont {A.~J.}\ \bibnamefont {Musser}},
  \bibinfo {author} {\bibfnamefont {K.}~\bibnamefont {Chen}}, \bibinfo {author}
  {\bibfnamefont {S.}~\bibnamefont {Athanasopoulos}}, \bibinfo {author}
  {\bibfnamefont {C.~K.}\ \bibnamefont {Yong}}, \bibinfo {author}
  {\bibfnamefont {Z.}~\bibnamefont {Zeng}}, \bibinfo {author} {\bibfnamefont
  {Q.}~\bibnamefont {Ye}}, \bibinfo {author} {\bibfnamefont {C.}~\bibnamefont
  {Chi}}, \bibinfo {author} {\bibfnamefont {J.~M.}\ \bibnamefont {Hodgkiss}},
  \bibinfo {author} {\bibfnamefont {J.}~\bibnamefont {Wu}}, \bibinfo {author}
  {\bibfnamefont {R.~H.}\ \bibnamefont {Friend}}, , \ and\ \bibinfo {author}
  {\bibfnamefont {N.~C.}\ \bibnamefont {Greenham}},\ }\href@noop {} {\bibfield
  {journal} {\bibinfo  {journal} {Adv. Funct. Mater.}\ }\textbf {\bibinfo
  {volume} {25}},\ \bibinfo {pages} {5452} (\bibinfo {year}
  {2015})}\BibitemShut {NoStop}%
\bibitem [{\citenamefont {Sanders}\ \emph
  {et~al.}(2016{\natexlab{a}})\citenamefont {Sanders}, \citenamefont
  {Kumarasamy}, \citenamefont {Pun}, \citenamefont {Steigerwald}, \citenamefont
  {Sfeir},\ and\ \citenamefont {Campos}}]{Sfeir16-1}%
  \BibitemOpen
  \bibfield  {author} {\bibinfo {author} {\bibfnamefont {S.~N.}\ \bibnamefont
  {Sanders}}, \bibinfo {author} {\bibfnamefont {E.}~\bibnamefont {Kumarasamy}},
  \bibinfo {author} {\bibfnamefont {A.~B.}\ \bibnamefont {Pun}}, \bibinfo
  {author} {\bibfnamefont {M.~L.}\ \bibnamefont {Steigerwald}}, \bibinfo
  {author} {\bibfnamefont {M.~Y.}\ \bibnamefont {Sfeir}}, \ and\ \bibinfo
  {author} {\bibfnamefont {L.~M.}\ \bibnamefont {Campos}},\ }\href@noop {}
  {\bibfield  {journal} {\bibinfo  {journal} {Angew. Chem.}\ }\textbf {\bibinfo
  {volume} {128}},\ \bibinfo {pages} {3434} (\bibinfo {year}
  {2016}{\natexlab{a}})}\BibitemShut {NoStop}%
\bibitem [{\citenamefont {Sanders}\ \emph
  {et~al.}(2016{\natexlab{b}})\citenamefont {Sanders}, \citenamefont
  {Kumarasamy}, \citenamefont {Pun}, \citenamefont {Appavoo}, \citenamefont
  {Steigerwald}, \citenamefont {Campos},\ and\ \citenamefont
  {Sfeir}}]{Sfeir16-2}%
  \BibitemOpen
  \bibfield  {author} {\bibinfo {author} {\bibfnamefont {S.~N.}\ \bibnamefont
  {Sanders}}, \bibinfo {author} {\bibfnamefont {E.}~\bibnamefont {Kumarasamy}},
  \bibinfo {author} {\bibfnamefont {A.~B.}\ \bibnamefont {Pun}}, \bibinfo
  {author} {\bibfnamefont {K.}~\bibnamefont {Appavoo}}, \bibinfo {author}
  {\bibfnamefont {M.~L.}\ \bibnamefont {Steigerwald}}, \bibinfo {author}
  {\bibfnamefont {L.~M.}\ \bibnamefont {Campos}}, \ and\ \bibinfo {author}
  {\bibfnamefont {M.~Y.}\ \bibnamefont {Sfeir}},\ }\href@noop {} {\bibfield
  {journal} {\bibinfo  {journal} {J. Am. Chem. Soc.}\ }\textbf {\bibinfo
  {volume} {138}},\ \bibinfo {pages} {7289} (\bibinfo {year}
  {2016}{\natexlab{b}})}\BibitemShut {NoStop}%
\bibitem [{\citenamefont {Sakuma}\ \emph {et~al.}(2016)\citenamefont {Sakuma},
  \citenamefont {Sakai}, \citenamefont {Araki}, \citenamefont {Mori},
  \citenamefont {Wada}, \citenamefont {Tkachenko},\ and\ \citenamefont
  {Hasobe}}]{Hasobe16}%
  \BibitemOpen
  \bibfield  {author} {\bibinfo {author} {\bibfnamefont {T.}~\bibnamefont
  {Sakuma}}, \bibinfo {author} {\bibfnamefont {H.}~\bibnamefont {Sakai}},
  \bibinfo {author} {\bibfnamefont {Y.}~\bibnamefont {Araki}}, \bibinfo
  {author} {\bibfnamefont {T.}~\bibnamefont {Mori}}, \bibinfo {author}
  {\bibfnamefont {T.}~\bibnamefont {Wada}}, \bibinfo {author} {\bibfnamefont
  {N.~V.}\ \bibnamefont {Tkachenko}}, \ and\ \bibinfo {author} {\bibfnamefont
  {T.}~\bibnamefont {Hasobe}},\ }\href@noop {} {\bibfield  {journal} {\bibinfo
  {journal} {J. Phys. Chem. A}\ }\textbf {\bibinfo {volume} {120}},\ \bibinfo
  {pages} {1867} (\bibinfo {year} {2016})}\BibitemShut {NoStop}%
\bibitem [{\citenamefont {Ito}\ \emph {et~al.}(2016)\citenamefont {Ito},
  \citenamefont {Nagami},\ and\ \citenamefont {Nakano}}]{Nakano16}%
  \BibitemOpen
  \bibfield  {author} {\bibinfo {author} {\bibfnamefont {S.}~\bibnamefont
  {Ito}}, \bibinfo {author} {\bibfnamefont {T.}~\bibnamefont {Nagami}}, \ and\
  \bibinfo {author} {\bibfnamefont {M.}~\bibnamefont {Nakano}},\ }\href@noop {}
  {\bibfield  {journal} {\bibinfo  {journal} {J. Phys. Chem. A}\ }\textbf
  {\bibinfo {volume} {120}},\ \bibinfo {pages} {6236} (\bibinfo {year}
  {2016})}\BibitemShut {NoStop}%
\bibitem [{\citenamefont {Zirzlmeier}\ \emph {et~al.}(2016)\citenamefont
  {Zirzlmeier}, \citenamefont {Casillas}, \citenamefont {Reddy}, \citenamefont
  {Coto}, \citenamefont {Lehnherr}, \citenamefont {Chernick}, \citenamefont
  {Papadopoulos}, \citenamefont {Thoss}, \citenamefont {Tykwinski},\ and\
  \citenamefont {Guldi}}]{Guldi16}%
  \BibitemOpen
  \bibfield  {author} {\bibinfo {author} {\bibfnamefont {J.}~\bibnamefont
  {Zirzlmeier}}, \bibinfo {author} {\bibfnamefont {R.}~\bibnamefont
  {Casillas}}, \bibinfo {author} {\bibfnamefont {S.~R.}\ \bibnamefont {Reddy}},
  \bibinfo {author} {\bibfnamefont {P.~B.}\ \bibnamefont {Coto}}, \bibinfo
  {author} {\bibfnamefont {D.}~\bibnamefont {Lehnherr}}, \bibinfo {author}
  {\bibfnamefont {E.~T.}\ \bibnamefont {Chernick}}, \bibinfo {author}
  {\bibfnamefont {I.}~\bibnamefont {Papadopoulos}}, \bibinfo {author}
  {\bibfnamefont {M.}~\bibnamefont {Thoss}}, \bibinfo {author} {\bibfnamefont
  {R.~R.}\ \bibnamefont {Tykwinski}}, \ and\ \bibinfo {author} {\bibfnamefont
  {D.~M.}\ \bibnamefont {Guldi}},\ }\href@noop {} {\bibfield  {journal}
  {\bibinfo  {journal} {Nanoscale}\ }\textbf {\bibinfo {volume} {8}},\ \bibinfo
  {pages} {10113} (\bibinfo {year} {2016})}\BibitemShut {NoStop}%
\bibitem [{\citenamefont {Lukman}\ \emph {et~al.}(2016)\citenamefont {Lukman},
  \citenamefont {Chen}, \citenamefont {Hodgkiss}, \citenamefont {Turban},
  \citenamefont {Hine}, \citenamefont {Dong}, \citenamefont {Wu}, \citenamefont
  {Greenham},\ and\ \citenamefont {Musser}}]{Musser16}%
  \BibitemOpen
  \bibfield  {author} {\bibinfo {author} {\bibfnamefont {S.}~\bibnamefont
  {Lukman}}, \bibinfo {author} {\bibfnamefont {K.}~\bibnamefont {Chen}},
  \bibinfo {author} {\bibfnamefont {J.~M.}\ \bibnamefont {Hodgkiss}}, \bibinfo
  {author} {\bibfnamefont {D.~H.~P.}\ \bibnamefont {Turban}}, \bibinfo {author}
  {\bibfnamefont {N.~D.~M.}\ \bibnamefont {Hine}}, \bibinfo {author}
  {\bibfnamefont {S.}~\bibnamefont {Dong}}, \bibinfo {author} {\bibfnamefont
  {J.}~\bibnamefont {Wu}}, \bibinfo {author} {\bibfnamefont {N.~C.}\
  \bibnamefont {Greenham}}, \ and\ \bibinfo {author} {\bibfnamefont {A.~J.}\
  \bibnamefont {Musser}},\ }\href@noop {} {\bibfield  {journal} {\bibinfo
  {journal} {Nat. Commun.}\ }\textbf {\bibinfo {volume} {7}},\ \bibinfo {pages}
  {13622} (\bibinfo {year} {2016})}\BibitemShut {NoStop}%
\bibitem [{\citenamefont {Sanders}\ \emph
  {et~al.}(2016{\natexlab{c}})\citenamefont {Sanders}, \citenamefont
  {Kumarasamy}, \citenamefont {Pun}, \citenamefont {Steigerwald}, \citenamefont
  {Sfeir},\ and\ \citenamefont {Campos}}]{Sfeir16-3}%
  \BibitemOpen
  \bibfield  {author} {\bibinfo {author} {\bibfnamefont {S.~N.}\ \bibnamefont
  {Sanders}}, \bibinfo {author} {\bibfnamefont {E.}~\bibnamefont {Kumarasamy}},
  \bibinfo {author} {\bibfnamefont {A.~B.}\ \bibnamefont {Pun}}, \bibinfo
  {author} {\bibfnamefont {M.~L.}\ \bibnamefont {Steigerwald}}, \bibinfo
  {author} {\bibfnamefont {M.~Y.}\ \bibnamefont {Sfeir}}, \ and\ \bibinfo
  {author} {\bibfnamefont {L.~M.}\ \bibnamefont {Campos}},\ }\href@noop {}
  {\bibfield  {journal} {\bibinfo  {journal} {Chem}\ }\textbf {\bibinfo
  {volume} {1}},\ \bibinfo {pages} {505} (\bibinfo {year}
  {2016}{\natexlab{c}})}\BibitemShut {NoStop}%
\bibitem [{\citenamefont {Johnson}\ \emph
  {et~al.}(2013{\natexlab{b}})\citenamefont {Johnson}, \citenamefont {Akdag},
  \citenamefont {Zamadar}, \citenamefont {Chen}, \citenamefont {Schwerin},
  \citenamefont {Paci}, \citenamefont {Smith}, \citenamefont {Havlas},
  \citenamefont {Miller}, \citenamefont {Ratner}, \citenamefont {Nozik},\ and\
  \citenamefont {Michl}}]{Michl13-3}%
  \BibitemOpen
  \bibfield  {author} {\bibinfo {author} {\bibfnamefont {J.~C.}\ \bibnamefont
  {Johnson}}, \bibinfo {author} {\bibfnamefont {A.}~\bibnamefont {Akdag}},
  \bibinfo {author} {\bibfnamefont {M.}~\bibnamefont {Zamadar}}, \bibinfo
  {author} {\bibfnamefont {X.}~\bibnamefont {Chen}}, \bibinfo {author}
  {\bibfnamefont {A.~F.}\ \bibnamefont {Schwerin}}, \bibinfo {author}
  {\bibfnamefont {I.}~\bibnamefont {Paci}}, \bibinfo {author} {\bibfnamefont
  {M.~B.}\ \bibnamefont {Smith}}, \bibinfo {author} {\bibfnamefont
  {Z.}~\bibnamefont {Havlas}}, \bibinfo {author} {\bibfnamefont {J.~R.}\
  \bibnamefont {Miller}}, \bibinfo {author} {\bibfnamefont {M.~A.}\
  \bibnamefont {Ratner}}, \bibinfo {author} {\bibfnamefont {A.~J.}\
  \bibnamefont {Nozik}}, \ and\ \bibinfo {author} {\bibfnamefont
  {J.}~\bibnamefont {Michl}},\ }\href@noop {} {\bibfield  {journal} {\bibinfo
  {journal} {J. Phys. Chem. B}\ }\textbf {\bibinfo {volume} {117}},\ \bibinfo
  {pages} {4680} (\bibinfo {year} {2013}{\natexlab{b}})}\BibitemShut {NoStop}%
\bibitem [{\citenamefont {Margulies}\ \emph {et~al.}(2016)\citenamefont
  {Margulies}, \citenamefont {Miller}, \citenamefont {Wu}, \citenamefont {Ma},
  \citenamefont {Schatz}, \citenamefont {Young},\ and\ \citenamefont
  {Wasielewski}}]{Wasielewski16}%
  \BibitemOpen
  \bibfield  {author} {\bibinfo {author} {\bibfnamefont {E.~A.}\ \bibnamefont
  {Margulies}}, \bibinfo {author} {\bibfnamefont {C.~E.}\ \bibnamefont
  {Miller}}, \bibinfo {author} {\bibfnamefont {Y.}~\bibnamefont {Wu}}, \bibinfo
  {author} {\bibfnamefont {L.}~\bibnamefont {Ma}}, \bibinfo {author}
  {\bibfnamefont {G.~C.}\ \bibnamefont {Schatz}}, \bibinfo {author}
  {\bibfnamefont {R.~M.}\ \bibnamefont {Young}}, \ and\ \bibinfo {author}
  {\bibfnamefont {M.~R.}\ \bibnamefont {Wasielewski}},\ }\href@noop {}
  {\bibfield  {journal} {\bibinfo  {journal} {Nat. Chem.}\ }\textbf {\bibinfo
  {volume} {8}},\ \bibinfo {pages} {1120} (\bibinfo {year} {2016})}\BibitemShut
  {NoStop}%
\bibitem [{\citenamefont {Varnavski}\ \emph {et~al.}(2015)\citenamefont
  {Varnavski}, \citenamefont {Abeyasinghe}, \citenamefont {Arag\'{o}},
  \citenamefont {{Serrano-P\'{e}rez}}, \citenamefont {Ort\'{i}}, \citenamefont
  {{L\'{o}pez Navarrete}}, \citenamefont {Takimiya}, \citenamefont {Casanova},
  \citenamefont {Casado},\ and\ \citenamefont {{Goodson, III}}}]{Goodson15}%
  \BibitemOpen
  \bibfield  {author} {\bibinfo {author} {\bibfnamefont {O.}~\bibnamefont
  {Varnavski}}, \bibinfo {author} {\bibfnamefont {N.}~\bibnamefont
  {Abeyasinghe}}, \bibinfo {author} {\bibfnamefont {J.}~\bibnamefont
  {Arag\'{o}}}, \bibinfo {author} {\bibfnamefont {J.~J.}\ \bibnamefont
  {{Serrano-P\'{e}rez}}}, \bibinfo {author} {\bibfnamefont {E.}~\bibnamefont
  {Ort\'{i}}}, \bibinfo {author} {\bibfnamefont {J.~T.}\ \bibnamefont
  {{L\'{o}pez Navarrete}}}, \bibinfo {author} {\bibfnamefont {K.}~\bibnamefont
  {Takimiya}}, \bibinfo {author} {\bibfnamefont {D.}~\bibnamefont {Casanova}},
  \bibinfo {author} {\bibfnamefont {J.}~\bibnamefont {Casado}}, \ and\ \bibinfo
  {author} {\bibfnamefont {T.}~\bibnamefont {{Goodson, III}}},\ }\href@noop {}
  {\bibfield  {journal} {\bibinfo  {journal} {J. Phys. Chem. Lett.}\ }\textbf
  {\bibinfo {volume} {6}},\ \bibinfo {pages} {1375} (\bibinfo {year}
  {2015})}\BibitemShut {NoStop}%
\bibitem [{\citenamefont {Chien}\ \emph {et~al.}(2015)\citenamefont {Chien},
  \citenamefont {Molina}, \citenamefont {Abeyasinghe}, \citenamefont
  {Varnavski}, \citenamefont {{Goodson, III}},\ and\ \citenamefont
  {Zimmerman}}]{Zimmerman15}%
  \BibitemOpen
  \bibfield  {author} {\bibinfo {author} {\bibfnamefont {A.~D.}\ \bibnamefont
  {Chien}}, \bibinfo {author} {\bibfnamefont {A.~R.}\ \bibnamefont {Molina}},
  \bibinfo {author} {\bibfnamefont {N.}~\bibnamefont {Abeyasinghe}}, \bibinfo
  {author} {\bibfnamefont {O.~P.}\ \bibnamefont {Varnavski}}, \bibinfo {author}
  {\bibfnamefont {T.}~\bibnamefont {{Goodson, III}}}, \ and\ \bibinfo {author}
  {\bibfnamefont {P.~M.}\ \bibnamefont {Zimmerman}},\ }\href@noop {} {\bibfield
   {journal} {\bibinfo  {journal} {J. Phys. Chem. C}\ }\textbf {\bibinfo
  {volume} {119}},\ \bibinfo {pages} {28258} (\bibinfo {year}
  {2015})}\BibitemShut {NoStop}%
\bibitem [{\citenamefont {Pariser}\ and\ \citenamefont
  {Parr}(1953)}]{pariser-parr}%
  \BibitemOpen
  \bibfield  {author} {\bibinfo {author} {\bibfnamefont {R.}~\bibnamefont
  {Pariser}}\ and\ \bibinfo {author} {\bibfnamefont {R.~G.}\ \bibnamefont
  {Parr}},\ }\href@noop {} {\bibfield  {journal} {\bibinfo  {journal} {J. Chem.
  Phys.}\ }\textbf {\bibinfo {volume} {21}},\ \bibinfo {pages} {466} (\bibinfo
  {year} {1953})}\BibitemShut {NoStop}%
\bibitem [{\citenamefont {Pople}(1953)}]{pople}%
  \BibitemOpen
  \bibfield  {author} {\bibinfo {author} {\bibfnamefont {J.~A.}\ \bibnamefont
  {Pople}},\ }\href@noop {} {\bibfield  {journal} {\bibinfo  {journal} {Trans.
  Faraday Soc.}\ }\textbf {\bibinfo {volume} {49}},\ \bibinfo {pages} {1375}
  (\bibinfo {year} {1953})}\BibitemShut {NoStop}%
\bibitem [{\citenamefont {Ohno}(1964)}]{ohno}%
  \BibitemOpen
  \bibfield  {author} {\bibinfo {author} {\bibfnamefont {K.}~\bibnamefont
  {Ohno}},\ }\href@noop {} {\bibfield  {journal} {\bibinfo  {journal} {Theor.
  Chim. Acta}\ }\textbf {\bibinfo {volume} {2}},\ \bibinfo {pages} {219}
  (\bibinfo {year} {1964})}\BibitemShut {NoStop}%
\bibitem [{\citenamefont {Klopman}(1964)}]{klopman}%
  \BibitemOpen
  \bibfield  {author} {\bibinfo {author} {\bibfnamefont {G.}~\bibnamefont
  {Klopman}},\ }\href@noop {} {\bibfield  {journal} {\bibinfo  {journal} {J.
  Am. Chem. Soc.}\ }\textbf {\bibinfo {volume} {86}},\ \bibinfo {pages} {4550}
  (\bibinfo {year} {1964})}\BibitemShut {NoStop}%
\bibitem [{\citenamefont {Race}\ \emph {et~al.}(2001)\citenamefont {Race},
  \citenamefont {Barford},\ and\ \citenamefont {Bursill}}]{race01}%
  \BibitemOpen
  \bibfield  {author} {\bibinfo {author} {\bibfnamefont {A.}~\bibnamefont
  {Race}}, \bibinfo {author} {\bibfnamefont {W.}~\bibnamefont {Barford}}, \
  and\ \bibinfo {author} {\bibfnamefont {R.~J.}\ \bibnamefont {Bursill}},\
  }\href@noop {} {\bibfield  {journal} {\bibinfo  {journal} {Phys. Rev. B}\
  }\textbf {\bibinfo {volume} {64}},\ \bibinfo {pages} {035208} (\bibinfo
  {year} {2001})}\BibitemShut {NoStop}%
\bibitem [{\citenamefont {Race}\ \emph {et~al.}(2003)\citenamefont {Race},
  \citenamefont {Barford},\ and\ \citenamefont {Bursill}}]{race03}%
  \BibitemOpen
  \bibfield  {author} {\bibinfo {author} {\bibfnamefont {A.}~\bibnamefont
  {Race}}, \bibinfo {author} {\bibfnamefont {W.}~\bibnamefont {Barford}}, \
  and\ \bibinfo {author} {\bibfnamefont {R.~J.}\ \bibnamefont {Bursill}},\
  }\href@noop {} {\bibfield  {journal} {\bibinfo  {journal} {Phys. Rev. B}\
  }\textbf {\bibinfo {volume} {67}},\ \bibinfo {pages} {245202} (\bibinfo
  {year} {2003})}\BibitemShut {NoStop}%
\bibitem [{\citenamefont {Soos}\ and\ \citenamefont
  {Ramasesha}(1984)}]{Soos84}%
  \BibitemOpen
  \bibfield  {author} {\bibinfo {author} {\bibfnamefont {Z.~G.}\ \bibnamefont
  {Soos}}\ and\ \bibinfo {author} {\bibfnamefont {S.}~\bibnamefont
  {Ramasesha}},\ }\href@noop {} {\bibfield  {journal} {\bibinfo  {journal}
  {Phys. Rev. B}\ }\textbf {\bibinfo {volume} {29}},\ \bibinfo {pages} {5410}
  (\bibinfo {year} {1984})}\BibitemShut {NoStop}%
\bibitem [{\citenamefont {Ramasesha}\ and\ \citenamefont
  {Soos}(1984)}]{Ramasesh84}%
  \BibitemOpen
  \bibfield  {author} {\bibinfo {author} {\bibfnamefont {S.}~\bibnamefont
  {Ramasesha}}\ and\ \bibinfo {author} {\bibfnamefont {Z.~G.}\ \bibnamefont
  {Soos}},\ }\href@noop {} {\bibfield  {journal} {\bibinfo  {journal} {Int. J.
  Quan. Chem.}\ }\textbf {\bibinfo {volume} {25}},\ \bibinfo {pages} {1003}
  (\bibinfo {year} {1984})}\BibitemShut {NoStop}%
\bibitem [{\citenamefont {Campbell}\ \emph {et~al.}(1990)\citenamefont
  {Campbell}, \citenamefont {Gammel},\ and\ \citenamefont {{Loh,
  Jr}}}]{Campbell90}%
  \BibitemOpen
  \bibfield  {author} {\bibinfo {author} {\bibfnamefont {D.~K.}\ \bibnamefont
  {Campbell}}, \bibinfo {author} {\bibfnamefont {J.~T.}\ \bibnamefont
  {Gammel}}, \ and\ \bibinfo {author} {\bibfnamefont {E.~Y.}\ \bibnamefont
  {{Loh, Jr}}},\ }\href@noop {} {\bibfield  {journal} {\bibinfo  {journal}
  {Phys. Rev. B}\ }\textbf {\bibinfo {volume} {42}},\ \bibinfo {pages} {475}
  (\bibinfo {year} {1990})}\BibitemShut {NoStop}%
\bibitem [{\citenamefont {Tandon}\ \emph {et~al.}(2003)\citenamefont {Tandon},
  \citenamefont {Ramasesha},\ and\ \citenamefont {Mazumdar}}]{Ramasesh03}%
  \BibitemOpen
  \bibfield  {author} {\bibinfo {author} {\bibfnamefont {K.}~\bibnamefont
  {Tandon}}, \bibinfo {author} {\bibfnamefont {S.}~\bibnamefont {Ramasesha}}, \
  and\ \bibinfo {author} {\bibfnamefont {S.}~\bibnamefont {Mazumdar}},\
  }\href@noop {} {\bibfield  {journal} {\bibinfo  {journal} {Phys. Rev. B}\
  }\textbf {\bibinfo {volume} {67}},\ \bibinfo {pages} {045109} (\bibinfo
  {year} {2003})}\BibitemShut {NoStop}%
\bibitem [{\citenamefont {Wolfsberg}\ and\ \citenamefont
  {Helmholz}(1952)}]{Wolfsberg52}%
  \BibitemOpen
  \bibfield  {author} {\bibinfo {author} {\bibfnamefont {M.}~\bibnamefont
  {Wolfsberg}}\ and\ \bibinfo {author} {\bibfnamefont {L.}~\bibnamefont
  {Helmholz}},\ }\href@noop {} {\bibfield  {journal} {\bibinfo  {journal} {J.
  Chem. Phys.}\ }\textbf {\bibinfo {volume} {20}},\ \bibinfo {pages} {837}
  (\bibinfo {year} {1952})}\BibitemShut {NoStop}%
\bibitem [{\citenamefont {Hoffmann}\ and\ \citenamefont
  {Lipscomb}(1962)}]{Hoffmann62}%
  \BibitemOpen
  \bibfield  {author} {\bibinfo {author} {\bibfnamefont {R.}~\bibnamefont
  {Hoffmann}}\ and\ \bibinfo {author} {\bibfnamefont {W.~N.}\ \bibnamefont
  {Lipscomb}},\ }\href@noop {} {\bibfield  {journal} {\bibinfo  {journal} {J.
  Chem. Phys.}\ }\textbf {\bibinfo {volume} {36}},\ \bibinfo {pages} {2179}
  (\bibinfo {year} {1962})}\BibitemShut {NoStop}%
\bibitem [{sup()}]{supple}%
  \BibitemOpen
  \href@noop {} {}\bibinfo {note} {See Supplemental Material at $[]$ for
  properties of low-lying singlet states of butadiene and hexatriene dimers,
  projection of $1{}^1B\otimes 1{}^1A$ and $T_1\otimes T_1$ with eigenstates of
  the dimer system within H{\"u}ckel and Hubbard model and time evolution
  profiles of unsubstituted and substituted butadiene and hexatriene
  dimers.}\BibitemShut {Stop}%
\bibitem [{\citenamefont {Iitaka}(1994)}]{Iitaka94}%
  \BibitemOpen
  \bibfield  {author} {\bibinfo {author} {\bibfnamefont {T.}~\bibnamefont
  {Iitaka}},\ }\href@noop {} {\bibfield  {journal} {\bibinfo  {journal} {Phys.
  Rev. E}\ }\textbf {\bibinfo {volume} {49}},\ \bibinfo {pages} {4684}
  (\bibinfo {year} {1994})}\BibitemShut {NoStop}%
\bibitem [{\citenamefont {Dutta}\ and\ \citenamefont
  {Ramasesha}(2007)}]{Dutta07}%
  \BibitemOpen
  \bibfield  {author} {\bibinfo {author} {\bibfnamefont {T.}~\bibnamefont
  {Dutta}}\ and\ \bibinfo {author} {\bibfnamefont {S.}~\bibnamefont
  {Ramasesha}},\ }\href@noop {} {\bibfield  {journal} {\bibinfo  {journal}
  {Computing Letters}\ }\textbf {\bibinfo {volume} {3}},\ \bibinfo {pages}
  {457} (\bibinfo {year} {2007})}\BibitemShut {NoStop}%
\bibitem [{\citenamefont {Chapra}\ and\ \citenamefont {Canale}(2012)}]{Chapra}%
  \BibitemOpen
  \bibfield  {author} {\bibinfo {author} {\bibfnamefont {S.~C.}\ \bibnamefont
  {Chapra}}\ and\ \bibinfo {author} {\bibfnamefont {R.~P.}\ \bibnamefont
  {Canale}},\ }\href@noop {} {\emph {\bibinfo {title} {Numerical Methods for
  Engineers}}},\ \bibinfo {edition} {6th}\ ed.\ (\bibinfo  {publisher} {McGraw
  Hill Education (India) Pvt. Ltd.},\ \bibinfo {address} {New Delhi},\ \bibinfo
  {year} {2012})\BibitemShut {NoStop}%
\bibitem [{\citenamefont {Crank}\ and\ \citenamefont
  {Nicolson}(1947)}]{Crank-Nicolson47}%
  \BibitemOpen
  \bibfield  {author} {\bibinfo {author} {\bibfnamefont {J.}~\bibnamefont
  {Crank}}\ and\ \bibinfo {author} {\bibfnamefont {P.}~\bibnamefont
  {Nicolson}},\ }\href@noop {} {\bibfield  {journal} {\bibinfo  {journal}
  {Mathematical Proceedings of the Cambridge Philosophical Society}\ }\textbf
  {\bibinfo {volume} {43}},\ \bibinfo {pages} {50} (\bibinfo {year}
  {1947})}\BibitemShut {NoStop}%
\bibitem [{\citenamefont {Rabi}(1937)}]{Rabi37}%
  \BibitemOpen
  \bibfield  {author} {\bibinfo {author} {\bibfnamefont {I.~I.}\ \bibnamefont
  {Rabi}},\ }\href@noop {} {\bibfield  {journal} {\bibinfo  {journal} {Phys.
  Rev.}\ }\textbf {\bibinfo {volume} {51}},\ \bibinfo {pages} {652} (\bibinfo
  {year} {1937})}\BibitemShut {NoStop}%
\bibitem [{\citenamefont {Tavan}\ and\ \citenamefont
  {Schulten}(1979{\natexlab{a}})}]{schulten79-1}%
  \BibitemOpen
  \bibfield  {author} {\bibinfo {author} {\bibfnamefont {P.}~\bibnamefont
  {Tavan}}\ and\ \bibinfo {author} {\bibfnamefont {K.}~\bibnamefont
  {Schulten}},\ }\href@noop {} {\bibfield  {journal} {\bibinfo  {journal} {J.
  Chem. Phys.}\ }\textbf {\bibinfo {volume} {70}},\ \bibinfo {pages} {5407}
  (\bibinfo {year} {1979}{\natexlab{a}})}\BibitemShut {NoStop}%
\bibitem [{\citenamefont {Tavan}\ and\ \citenamefont
  {Schulten}(1979{\natexlab{b}})}]{schulten79-2}%
  \BibitemOpen
  \bibfield  {author} {\bibinfo {author} {\bibfnamefont {P.}~\bibnamefont
  {Tavan}}\ and\ \bibinfo {author} {\bibfnamefont {K.}~\bibnamefont
  {Schulten}},\ }\href@noop {} {\bibfield  {journal} {\bibinfo  {journal} {J.
  Chem. Phys.}\ }\textbf {\bibinfo {volume} {70}},\ \bibinfo {pages} {5414}
  (\bibinfo {year} {1979}{\natexlab{b}})}\BibitemShut {NoStop}%
\bibitem [{\citenamefont {Tavan}\ and\ \citenamefont
  {Schulten}(1986)}]{schulten86}%
  \BibitemOpen
  \bibfield  {author} {\bibinfo {author} {\bibfnamefont {P.}~\bibnamefont
  {Tavan}}\ and\ \bibinfo {author} {\bibfnamefont {K.}~\bibnamefont
  {Schulten}},\ }\href@noop {} {\bibfield  {journal} {\bibinfo  {journal} {J.
  Chem. Phys.}\ }\textbf {\bibinfo {volume} {85}},\ \bibinfo {pages} {6602}
  (\bibinfo {year} {1986})}\BibitemShut {NoStop}%
\bibitem [{\citenamefont {Soos}\ \emph {et~al.}(1993)\citenamefont {Soos},
  \citenamefont {Ramasesha},\ and\ \citenamefont {{Galv\~{a}o}}}]{Soos93}%
  \BibitemOpen
  \bibfield  {author} {\bibinfo {author} {\bibfnamefont {Z.~G.}\ \bibnamefont
  {Soos}}, \bibinfo {author} {\bibfnamefont {S.}~\bibnamefont {Ramasesha}}, \
  and\ \bibinfo {author} {\bibfnamefont {D.~S.}\ \bibnamefont {{Galv\~{a}o}}},\
  }\href@noop {} {\bibfield  {journal} {\bibinfo  {journal} {Phys. Rev. Lett.}\
  }\textbf {\bibinfo {volume} {71}},\ \bibinfo {pages} {1609} (\bibinfo {year}
  {1993})}\BibitemShut {NoStop}%
\bibitem [{\citenamefont {Baeriswyl}\ \emph {et~al.}(1992)\citenamefont
  {Baeriswyl}, \citenamefont {Campbell},\ and\ \citenamefont
  {Mazumdar}}]{Baeriswyl}%
  \BibitemOpen
  \bibfield  {author} {\bibinfo {author} {\bibfnamefont {D.}~\bibnamefont
  {Baeriswyl}}, \bibinfo {author} {\bibfnamefont {D.~K.}\ \bibnamefont
  {Campbell}}, \ and\ \bibinfo {author} {\bibfnamefont {S.}~\bibnamefont
  {Mazumdar}},\ }\href@noop {} {\emph {\bibinfo {title} {Conjugated Conducting
  Polymers}}},\ edited by\ \bibinfo {editor} {\bibfnamefont {H.}~\bibnamefont
  {Kiess}},\ \bibinfo {series} {Springer Series in Solid-State Sciences}, Vol.\
  \bibinfo {volume} {102}\ (\bibinfo  {publisher} {Springer},\ \bibinfo
  {address} {Berlin},\ \bibinfo {year} {1992})\BibitemShut {NoStop}%
\bibitem [{\citenamefont {Shuai}\ \emph {et~al.}(1997)\citenamefont {Shuai},
  \citenamefont {Br\'{e}das}, \citenamefont {Pati},\ and\ \citenamefont
  {Ramasesha}}]{Shuai97}%
  \BibitemOpen
  \bibfield  {author} {\bibinfo {author} {\bibfnamefont {Z.}~\bibnamefont
  {Shuai}}, \bibinfo {author} {\bibfnamefont {J.~L.}\ \bibnamefont
  {Br\'{e}das}}, \bibinfo {author} {\bibfnamefont {S.~K.}\ \bibnamefont
  {Pati}}, \ and\ \bibinfo {author} {\bibfnamefont {S.}~\bibnamefont
  {Ramasesha}},\ }\href@noop {} {\bibfield  {journal} {\bibinfo  {journal}
  {Phys. Rev. B}\ }\textbf {\bibinfo {volume} {56}},\ \bibinfo {pages} {9298}
  (\bibinfo {year} {1997})}\BibitemShut {NoStop}%
\bibitem [{\citenamefont {Ren}\ \emph {et~al.}(2017)\citenamefont {Ren},
  \citenamefont {Peng}, \citenamefont {Zhang}, \citenamefont {Yi},\ and\
  \citenamefont {Shuai}}]{Shuai17}%
  \BibitemOpen
  \bibfield  {author} {\bibinfo {author} {\bibfnamefont {J.}~\bibnamefont
  {Ren}}, \bibinfo {author} {\bibfnamefont {Q.}~\bibnamefont {Peng}}, \bibinfo
  {author} {\bibfnamefont {X.}~\bibnamefont {Zhang}}, \bibinfo {author}
  {\bibfnamefont {Y.}~\bibnamefont {Yi}}, \ and\ \bibinfo {author}
  {\bibfnamefont {Z.}~\bibnamefont {Shuai}},\ }\href@noop {} {\bibfield
  {journal} {\bibinfo  {journal} {J. Phys. Chem. Lett.}\ }\textbf {\bibinfo
  {volume} {8}},\ \bibinfo {pages} {2175} (\bibinfo {year} {2017})}\BibitemShut
  {NoStop}%
\end{thebibliography}%

\end{document}